\documentclass[a4paper,10pt,twocolumn]{article}

\usepackage{amsmath,amssymb}
\usepackage{graphicx}
\usepackage{natbib}

\newcommand{\Nl}{\,\mathrm{N}}
\newcommand{\Wl}{\,\mathrm{W}}

\newcommand{\mean}[1]{\left<#1\right>}

\newcommand{\Dtr}{{\mathrm{\Delta}_r\theta}}
\newcommand{\Dtz}{{\mathrm{\Delta}_z\theta}}
\newcommand{\muq}[2]{\mu_{#2}\left(#1\right)}
\newcommand{\s}{\nobreak\mbox{$\;$s}}
\newcommand{\mn}{\nobreak\mbox{$\;$min}}
\newcommand{\hrs}{\nobreak\mbox{$\;$hour}}
\newcommand{\m}{\nobreak\mbox{$\;$m}}
\newcommand{\kg}{\nobreak\mbox{$\;$kg}}

\newcommand{\dC}{\nobreak\mbox{$\;^\circ$C}}
\newcommand{\Hz}{\nobreak\mbox{$\;$Hz}}

\begin{document}

\title{Temperature statistics above a deep-ocean sloping boundary.}

\author{Andrea A. Cimatoribus\thanks{Email address for correspondence: \mbox{Andrea.Cimatoribus@nioz.nl}}, H. van Haren\\
\small Royal Netherlands Institute for Sea Research, \\
\small Landsdiep 4, 1797SZ, 't Horntje, NH, the Netherlands}
\date{\today}

\maketitle

\begin{abstract}
We present a detailed analysis of the temperature statistics in an oceanographic observational dataset.
The data are collected using a moored array of thermistors, $100\m$ tall and starting $5\m$ above the bottom, deployed during four months above the slopes of a Seamount in the north-eastern Atlantic Ocean.
Turbulence at this location is strongly affected by the semidiurnal tidal wave.
Mean stratification is stable in the entire dataset.
We compute structure functions, of order up to 10, of the distributions of temperature increments.
Strong intermittency is observed, in particular, during the downslope phase of the tide, and farther from the solid bottom.
In the lower half of the mooring during the upslope phase, the temperature statistics are consistent with those of a passive scalar.
In the upper half of the mooring, the temperature statistics deviate from those of a passive scalar, and evidence of turbulent convective activity is found.
The downslope phase is generally thought to be more shear-dominated, but our results suggest on the other hand that convective activity is present.
High-order moments also show that the turbulence scaling behaviour breaks at a well-defined scale (of the order of the buoyancy length scale), which is however dependent on the flow state (tidal phase, height above the bottom).
At larger scales, wave motions are dominant.
We suggest that our results could provide an important reference for laboratory and numerical studies of mixing in geophysical flows.
\end{abstract}

\section{Introduction}
\label{sec:introduction}

Turbulence above sloping ocean boundaries has been the subject of several studies since the work of \citet{munk_abyssal_1966}, who hypothesised that enhanced turbulence close to the solid boundary plays an important role in maintaining the ocean stratified in density from the surface to the bottom.
The dynamics of the benthic boundary layer above a slope has been the subject of several studies \citep[e.g.][]{garrett_role_1990,thorpe_variability_1990,garrett_marginal_1991,van_haren_measurements_1994,alford_observations_2000,slinn_modeling_2003,gayen_boundary_2011,gayen_negative_2011}.
Turbulence close to sloping boundaries is enhanced by the breaking of internal waves \citep[see the review of][]{lamb_internal_2014} which, from a broader point of view, dissipates tidal energy \citep{munk_abyssal_1998} and, to a lesser extent, eddy kinetic energy \citep{zhai_significant_2010}.

Here, we consider turbulence above a deep-ocean sloping boundary from the particular perspective of scalar (temperature) statistics.
The interest in the scalar statistics embedded in a turbulent flow is manifold.
The scalar field is highly intermittent, and its statistical moments bear the marks of this intermittency.
In relation to this, the study of the statistics of a scalar is the basis for understanding the dynamics driving its mixing, and for reliably estimating and modelling mixing itself.
The statistics of scalar tracers in a turbulent flow is a widely studied topic in the laboratory as well as numerically and analytically \citep[see the reviews of][]{shraiman_scalar_2000,warhaft_passive_2000}; most often the focus is on grid-generated turbulence.

On the other hand, intermittency of scalar fields in strongly stratified turbulent flows, where the buoyancy force is expected to be dominant, has received limited attention \citep[e.g.][using numerical simulations]{brethouwer_passive_2008}.
While we are unaware of laboratory studies on this topic, \citet{seuront_multifractal_1999} discussed temperature fluctuations in the surface ocean boundary layer in terms of multifractal intermittency.
\citet{frola_experimental_2014} recently reported observations of the probability density function (pdf) of temperature increments in the atmospheric boundary layer, discussing it in terms of a bifractal intermittency model.

We must stress that observations in the ocean and atmosphere often include some temperature statistics, but these are virtually always limited to second order, i.e.~to spectra \citep[see, for instance,][and references therein]{riley_stratified_2008}.
The oceanographic literature usually considers observations of temperature at the dissipation scale with the aim of estimating turbulent diffusivity and dissipation rate by using the \citet{osborn_oceanic_1972} model, and only second-order statistics are presented.
An exception is \citet{thorpe_skewness_1991}, where a possible link between the skewness of temperature increments and convection in benthic boundary layers was discussed.
Possibly more often, ocean temperature (or density) observations are collected at lower resolution, comparable to the Ozmidov scale \citep[e.g.][]{alford_observations_2000}.

This work presents a detailed study of the temperature statistics above a deep-ocean sloping boundary, focusing on the intermittency properties of the scalar.
The results are discussed comparing them to laboratory experiments with both passive (shear turbulence) and active (convective turbulence) scalars.
By necessity, our observations provide an average view of fluid regions having different Reynolds number, different mean shear and different mean stratification.
While this is to some extent a limitation, it should be kept in mind that geophysical turbulence is intrinsically sporadic i.e.~intermittent also at scales larger than those of the inertial range \citep{rorai_turbulence_2014}.
This fact must be taken into account when studying the scalar statistics in a stratified flow, and is naturally included in real-ocean observations.
Furthermore, laboratory and numerical experiments mainly focus on grid-generated turbulence or on the evolution of a particular kind of instability, usually Kelvin--Helmholtz. 
A unique forcing scale (grid length, most unstable wavenumber) can usually be identified.
In the present observations, on the other hand, there is no obvious separation between breaking waves (the main forcing) and turbulence.
Instabilities such as those produced in the laboratory are seldom recognised in the ocean, even if new observations may change this \citep[][and references therein]{van_haren_deep-ocean_2010,haren_stratified_2013,smyth_ocean_2012}.
For this reason, the relevance of the conclusions drawn from the study of Kelvin--Helmholtz instabilities for the real ocean is still not clear \citep{thorpe_kelvinhelmholtz_2012}.
In this context, the main aim of this work is to help bridge the gap between laboratory and geophysical flows, by analysing the observations in a way that is familiar to the fluid dynamics community (pdf's, moments, structure functions), and thus could serve as a reference.

The data used in this work have been collected using a moored array of thermistors deployed above the slopes of Seamount Josephine, in the north-eastern Atlantic Ocean.
Motions at this location are dominated by the tide.
The thermistors used, built in-house at the Royal Netherlands Institute for Sea Research (NIOZ), measure temperature variations at a relatively high temporal resolution ($1\s$) for long periods of time (several months).
While their temporal resolution is generally insufficient to resolve the complete inertial range of turbulence (nor to resolve the dissipation scale), they do capture part of the turbulence cascade.
In this work, we concentrate on time-averaged statistics, distinguishing between the upslope and the downslope phases of the tide, and distinguishing different parts of the moored array, i.e.~different depth ranges.
On the other hand, we do not consider time modulation of turbulence properties other than the tidal one, i.e.~we do not consider the modulation by inertial waves, mesoscale eddies or other low-frequency processes.
Time averages are a necessary starting point, enabling better-converged statistics, and thus enabling higher-order statistics to be reliably computed.

We will consider almost exclusively fluctuations and increments of temperature within the horizontal plane.
This is motivated by the fact that the resolution of our data is substantially better in the horizontal direction ($0.2\m$ by using Taylor's hypothesis; see section \ref{sec:velocity}) than in the vertical ($0.7\m$).
The difference in resolution is particularly relevant if the anisotropy of turbulence in a strongly stratified environment is taken into account, with length scales much smaller along the vertical direction (approximately parallel to the mean density gradient) than in the horizontal (approximately perpendicular to the mean density gradient) \cite{billant_self-similarity_2001}.

The paper is organised as follows. 
The temperature data used throughout the paper are described in section \ref{sec:data}.
Velocity observations are briefly described in section \ref{sec:velocity}.
The statistical observables considered are concisely presented in section \ref{sec:stats}.
The results are presented in section \ref{sec:results}; discussion and conclusions follow in section \ref{sec:conclusions}.

\section{Methods}
\label{sec:Methods}

\subsection{Temperature data}
\label{sec:data}

The data used in this work are obtained from a mooring deployed between spring and late summer 2013 in the north-eastern Atlantic Ocean on the slopes of Seamount Josephine (see table \ref{tab:mooring} for details).
The mooring has 144 ``NIOZ4'' thermistors \citep[an evolved version of the ones described in][]{van_haren_nioz3:_2009} taped on a nylon-coated steel cable at intervals of $0.7\m$.
The thermistors sampled temperature at a rate of $1\Hz$ with a precision better than $1 \,\mathrm{mK}$.
The time series from each thermistor is approximately $10^7\s$ long, with the entire dataset containing approximately $10^9$ measurements.
The cable was attached to a $500\kg$ ballast weight at the bottom, and to an elliptical buoy at the top.
The high net buoyancy (approximately $300\kg$) together with the small size of the sensors, limiting the drag on the cable, guarantee that measurements are very close to Eulerian, as checked also in similar previous moorings \citep{van_haren_high-resolution_2009,van_haren_detailed_2012}.
The pressure and tilt sensors of the acoustic Doppler current profiler (ADCP) mounted in the top buoy, $210\m$ above the bottom, recorded movements smaller than $0.2\m$ in the vertical direction and $3\m$ in the horizontal one. 
All the thermistors performed satisfactorily for the whole deployment, with a noise level of approximately $5 \times 10^{-5}\dC$.
Calibration is applied to the raw data and the drift in the response of the thermistor electronics, visible over periods longer than a few weeks, is compensated for.

At the mooring location, temperature and velocity variance are dominated by the semidiurnal tidal period (in particular by the $M_2$ lunar period of $12.42\hrs$), as shown by the velocity spectrum in figure \ref{fig:velocity} (left panel).
Tidal variations are visible in the temperature record as successive cooling and warming trends.
Throughout this paper, we consider various quantities computed separately for the cooling and warming tidal phases.
For simplicity, we will usually refer to the cooling phase as ``upslope'' tide and to the warming phase as ``downslope'' tide.
The two phases are considered separately because statistical properties vary markedly between the two.
The upslope and downslope phases are defined by the sign of the time derivative (negative and positive, respectively), of the vertically averaged temperature signal, band pass-filtered at the semidiurnal frequency.
This definition produces two subsets of the data that have virtually the same size.
Two snapshots are shown in figure \ref{fig:temperature}.
The two panels show approximately $30\mn$ of data each, during downslope (labelled ``down'') and upslope (labelled ``up'') tidal phases, respectively.
Figure \ref{fig:temperature} also shows the four vertical segments of the mooring (A, B, C, D) considered separately in the rest of the work, each approximately $25\m$ high and including 36 thermistors (table \ref{tab:segments}).
From figure \ref{fig:temperature}, we can see that, during both the upslope and the downslope phases, water is stratified at the mooring location.
This is confirmed by table \ref{tab:stratification}, reporting mean, first and ninety-ninth percentiles of the buoyancy frequency ($N$) in the four segments of the mooring and during the two tidal phases separately.
Stratification is computed from $25\s$ averages of temperature, with profiles vertically reordered to obtain a statically stable stratification\protect\footnote{This procedure, used to obtain an estimate of the background stratification, removes all instabilities from the vertical profiles. On the other hand, unstratified regions are not removed from profiles and would still produce vanishing values of the buoyancy frequency, if present. See \citet{winters_diascalar_1996} for a formal justification of the reordering procedure in the context of mixing estimation.}.
This observation is consistent with those by \citet{van_haren_detailed_2012}, which extended down to within $0.5\m$ from the solid boundary, differently from those discussed here, having the deepest thermistor $5\m$ above the bottom.
More details on the computation of buoyancy frequency using these data can be found in \citet{cimatoribus_comparison_2014}.
The mean buoyancy frequency is slightly higher during the upslope tidal phase, with the exception of the lowest segment, segment A.

\subsection{Velocity data and Taylor's hypothesis}
\label{sec:velocity}

A Teledyne/RDI 75 kHz ADCP was mounted in the top buoy of the mooring, $210\m$ above the bottom.
The ADCP was pointed downwards, resolving current and echo intensity observations every $5 \m$ vertically, between $15$ and $185\m$ above the bottom.
Horizontal vector velocity is measured, but only the magnitude within the depth range of temperature measurements is considered here.
We will refer to the magnitude of velocity simply as ``velocity'' in the following, unless stated otherwise.
The sampling interval was $600 \s$.

Despite the limited resolution, velocity data can be used to transform the measured temperature from the temporal to the spatial domain, according to the Taylor's hypothesis of frozen turbulence.
Here, we must take into account that the mean velocity is not constant, following the approach of \citet{pinton_correction_1994}.
The velocity field $v$ is first decomposed into a mean component $\mean{v}$ and a fluctuating one $v'$ using a low-pass Butterworth filter with stop band $1/\sigma=3000\s$.
On average, this can be considered the forcing frequency, based on the frequency spectra of temperature fluctuations (not shown).
Approximately at this value, the spectral slope decreases moving towards higher frequencies, approaching the value of $-5/3$ typical of the inertial range of turbulence in both homogeneous and stratified environments \citep{tennekes_first_1972,lindborg_vertical_2009}.
A similar transition is observed in the velocity spectrum in figure \ref{fig:velocity}, but in this case the turbulence range is mostly undetected and contaminated by noise (light blue line in the figure).
The pdf's of $v$, $\mean{v}$ and $v'$ are shown in figure \ref{fig:velocity} (right panel).
Mean velocity is higher during the upslope phase of the tide, while the pdf's of velocity fluctuations are similar for the upslope and downslope phases.

In order to transform the temperature time series of each thermistor into the spatial domain, the mean velocity is averaged, along the vertical coordinate, within each of the four segments of the mooring.
Since velocity is not constant in time, the spatial series thus obtained has a non-constant sampling rate.
Linear interpolation is performed to a constant time step of $0.2\m$, which reduces the size of the dataset by approximately a factor of 2.

The results of this transformation have to be evaluated with care, since we cannot assess the goodness of the Taylor's hypothesis, lacking information on the turbulent velocity.
This is particularly true closer to the bottom, where turbulence and shear are stronger.
Turbulence intensity ($\mean{v'^2}^{1/2}/\mean{v}$) computed from the ADCP data is much smaller than unity most of the time.
While this gives support for the use of Taylor's hypothesis, we do not emphasise this result since the ADCP resolution is insufficient (both in time and in the vertical) to detect most of the turbulence range, and noise contaminates the highest frequency range of ADCP measurements\protect\footnote{The presence of noise artificially increases the estimated turbulence intensity.}.
In the next subsections, the results are mostly based on the spatial (Taylor-transformed) data.
The same quantities reported here have been computed also on the original (time series) data, with qualitatively similar results.
Relatively short spatial separations are considered, which correspond in the time domain to time intervals no longer than approximately one hour.

\subsection{Statistical description}
\label{sec:stats}

A power spectrum can fully characterise a dataset only if the distribution of the data is approximately Gaussian.
This is usually not the case for temperature (and scalars in general) in a turbulent flow, nor is it the case here.
For this reason, our description of the dataset is extended to higher-order moments.
In doing so, we follow the example of the analyses of laboratory data available in the literature, mostly concerning passive scalars in grid-generated turbulence; see for example the works by \citet{thoroddsen_exponential_1992} and by \citet{mydlarski_passive_1998}, and  the reviews by \citet{warhaft_passive_2000} and \citet{shraiman_scalar_2000}.
We also make use of some of the concepts developed in \citet{zhou_plume_2002} for convectively generated turbulence.
We thus apply the now-classic methods of these works to the case of a strongly stratified ocean environment, with the aim of contributing to the understanding of the latter.
In fact, we find that the results from controlled environments are often comparable to those from our observational data.

We define here a common notation for the normalised moments of a distribution, defined for a generic variable $x$ at order $q$ by
\begin{equation}
  \muq{x}{q} = \frac{\mean{\left(x-\mean{x}\right)^q}}{\mean{\left(x-\mean{x}\right)^2}^{\frac{q}{2}}},
  \label{eq:norm_mom}
\end{equation}
with the average indicated by the angular brackets.
With this notation, the third-order moment of $x$, that is the skewness, is written as $\muq{x}{3}$ and the fourth-order moment, the flatness, as $\muq{x}{4}$, and both are defined by equation \ref{eq:norm_mom}.
These two moments have received particular attention in the literature \citep{warhaft_passive_2000}, as a way to characterise intermittency and anisotropy (skewness).

In section \ref{sec:fluctuations}, we consider the temperature fluctuations, $\theta$, using the full dataset in the time domain (not using Taylor's hypothesis).
Fluctuations are defined as the deviation from the linear trend computed during a single tidal phase ($6\hrs$ interval).
We compute the skewness and flatness of the fluctuations for each thermistor in order to assess the dependence on the height above the bottom.
Results from upslope and downslope tidal phases are averaged separately, after normalisation by the standard deviation of the single tidal phase.

Using the full dataset, skewness and flatness are determined by a combination of waves and turbulence that characterises a strongly stratified environment.
In particular, there is a strong influence from the waves at periods shorter than the tidal one, not removed by the linear detrending performed before computing the moments of the distribution.
In an attempt to separate turbulence from waves, the data can be filtered.
However, there is no obvious scale separation between waves and turbulence; this distinction may in fact be ill-defined.
The fluctuations seem to be a continuum of waves of different degree of nonlinearity at frequencies lower than the (local) buoyancy frequency and turbulent motions at frequencies both below and, mainly, above $N$ \citep[see e.g.][]{billant_self-similarity_2001}.
Keeping in mind that the filtering cannot separate, or define, these components, it still proves to be an instructive exercise, discussed in section \ref{sec:fluctuations}.
We use a Butterworth filter removing frequencies below $1/5600\s^{-1}$, a frequency approximately equal to the lowest values of the buoyancy frequency (see table \ref{tab:stratification}); qualitatively similar results are obtained using a higher stop-band frequency).

If the quantity $x$ in equation \ref{eq:norm_mom} is an increment, whose value depends on the distance between two measurements (in either time or space), then $\muq{x}{q}$ depends on the same distance too, and is usually called a structure function.
In particular, in section \ref{sec:sk_and_fl} we consider the temperature increments:
\begin{equation}
    \Dtr=\theta(r_0+r)-\theta(r_0),
    \label{eq:incrm}
\end{equation}
as a function of the separation distance $r$ (assuming homogeneity), usually along the horizontal direction (i.e.~using spatial data exploiting Taylor's hypothesis).
Temperature increments are particularly useful when dealing with non-stationary time series \citep{davis_multifractal_1994}. 
Increments enable one to naturally study turbulence and waves together without the need for filtering.

Further insight into the turbulence dynamics can be gained by studying the behaviour of generalised structure functions of temperature increments \citep{warhaft_passive_2000}
\begin{equation}
    \gamma_q\equiv\gamma_q(r)=\mean{\left|\Dtr\right|^q}.
    \label{eq:gsf}
\end{equation}
This is done in section \ref{sec:saturation}.
Generalised structure functions are a standard tool to characterise the intermittency of the velocity and scalar fields in a turbulent flow \citep{frisch_turbulence_1996,shraiman_scalar_2000,warhaft_passive_2000}.
Here, we obviously focus on the intermittency of the scalar (temperature).
According to the non-intermittent theory of Kolmogorov--Obukhov--Corrsin, the temperature field is expected to be fully self-similar within the inertial range, resulting in a structure function $\gamma_q$ having a scaling behaviour of the form \citep{warhaft_passive_2000}
\begin{equation}
    \gamma_q \sim r^{\zeta(q)}\mathrm{, with\; } \zeta(q) = q/3,
    \label{eq:zeta}
\end{equation}
for every order $q$ \citep[see also the discussion of this in][]{frola_experimental_2014}.
Note that a self-similar behaviour is compatible with a Gaussian distribution, but does not imply it.
For $q=2$, the structure function embeds the same information as the spectrum, with the slope $\zeta(2)=2/3$ of the structure function being the Fourier transform of the $-5/3$ slope of the fluctuations in wavenumber space.
While for $q=2$ the prediction of the theory is usually close to the values measured in the laboratory, for $q>2$ there is an increasing discrepancy between the predicted $q/3$ and the observed $\zeta(q)$, the latter being systematically below the theoretical value, possibly reaching an asymptotic value $\zeta_\infty$, i.e.~saturating, for sufficiently high orders ($q\approx 10$).
This saturation is related to the presence of ramp--cliff structures in the scalar field, in connection with an accumulation of sharp gradients at small scales.
A value of $\zeta(q)$ lower than expected is the consequence of an excess of these large gradients at small scales (the cliff structures) and of a lack of large increments at large scales (the ramp structures).
Evidence for saturation of moments has been reported in both passive scalars (shear-dominated turbulence) and active scalars (convectively driven turbulence).
However, the estimated asymptotic value of $\zeta_\infty$ is different for the two cases.
In particular, values of $\zeta_\infty\approx 1.4$ are reported for passive scalars in grid turbulence \citep{warhaft_passive_2000,celani_fronts_2001}, while smaller values ($\zeta_\infty\approx 0.8$) are observed in convectively driven turbulence \citep{zhou_plume_2002,celani_scaling_2002}.

In section \ref{sec:plus_and_minus}, we will consider ``plus'' and ``minus'' increments, introduced by \citet{zhou_plume_2002} in the context of convectively driven scalar turbulence, and used here to better investigate the passive/active nature of temperature.
They are defined as $\mathrm{\Delta}_\tau\theta^{\pm}=\left( \left| \mathrm{\Delta}_\tau\theta \right|\pm \mathrm{\Delta}_\tau\theta \right)/2$, and simply separate positive from negative increments in time (subscript $\tau$), measured at a fixed position.
A plus (minus) increment is zero for all negative (positive) increments, otherwise it is equal to the absolute value of the increment.
A hot plume rising through a thermistor produces a steep increase in temperature with its leading cap, but a relatively gentle temperature decrease with its tail.
This asymmetry between positive and negative temperature increments should show up in the skewness of the plus and minus increment distributions, with higher skewness for the plus increments (note that plus and minus increments are positive definite).
The laboratory results of \citet{zhou_plume_2002} indeed show this asymmetry in the convective region of a Rayleigh--B\'enard cell, for increments at inertial and dissipation separation distances.
More generally, here we use plus and minus increments to study if ramp--cliff structures have a preferential orientation.
In our case, convective plumes, if present, are advected through the thermistor array according to Taylor's hypothesis, rather than vertically rising through the array.
Consequently, we have to consider spatial increments of temperature rather than time increments, and we do so along both the horizontal and along the vertical directions.
Plus and minus increments are thus defined here as $\Dtr^{\pm}=\left( \left| \Dtr \right|\pm \Dtr \right)/2$ for increments in the horizontal direction, and similarly as $\Dtz^{\pm}=\left( \left| \Dtz \right|\pm \Dtz \right)/2$ along the vertical.
Here, $\Dtz$ is defined by equation \ref{eq:incrm}, with $r$ substituted by $z$, the upward vertical coordinate.
Given the relatively small number of sensors within each segment and the relatively low vertical resolution, $\Dtz^\pm$ is computed only at the thermistor spacing, i.e.~$0.7\m$.

\section{Results}
\label{sec:results}

Velocities are higher during the upslope phase of the tide (see figure \ref{fig:velocity}, right panel, black triangles).
Deviations of velocity from the vertical average explain approximately 25\% of the variance.
Shear is computed at the $5\m$ vertical resolution of the ADCP, as the average of the magnitude of the vertical gradient of the horizontal (vector) velocity.
Shear is similar during the two tidal phases, approximately equal to $3.2\times10^{-3}\s^{-1}$ (as a mean of the instantaneous shear). 
Mean velocity is virtually constant in the upper half of the mooring, and decreases by approximately 20\% in the lower half.
The temporal and spatial resolution of the ADCP is such that turbulent motions cannot be studied in any detail.
Velocity has a strong component along the local isobath during both tidal phases.

Figure \ref{fig:temperature} hints at some important differences between the two tidal phases.
During the downslope phase, the sharp interface above the cold near-bottom layer is lower and thinner.
Overturns at the sharp interface are also smaller than those observed during the upslope phase.
Both these elements suggest that turbulence is weaker during the downslope phase of the tide, as already observed by \citet{van_haren_detailed_2012} and also by \citet{moum_convectively_2004} in a shallower environment.
This fact is confirmed by calculations on this dataset using similar methods to \citet{van_haren_detailed_2012} (not shown).
A peculiar feature of both panels in figure \ref{fig:temperature} is that some of the turbulent structures, in particular in the upper half of the mooring, remind one of convective plumes.
Particularly striking for its size is the mushroom-like plume during the upslope phase (black arrow in the figure).

\subsection{Temperature horizontal wavenumber spectra}
\label{sec:spectra}

As a starting point in the statistical characterisation of the temperature field, horizontal wavenumber spectra of temperature fluctuations compensated by $k^{-5/3}$ are shown in figure \ref{fig:spectra}.
The figure shows spectra averaged during the two tidal phases (upslope, downslope) and in each mooring segment.

During the upslope tide (upper four spectra in figure \ref{fig:spectra}), an inertial range is clearly visible in all mooring segments as a horizontal slope, approximately in the $k^{-1}$ range from $5$ to $100\m$.
For segments A and B, there is also a hint of a local minimum at $k^{-1}\approx 100\m$, possibly the sign of a separation between the gravity-wave spectrum (lower wavenumbers, higher slope) and the turbulence band.
High-wavenumber roll-off of the spectrum is visible above $k^{-1}\approx5\m$, but the roll-off should be interpreted with care, since the $1\Hz$ sampling rate of the thermistors is insufficient to resolve the dissipation scales (of both momentum and temperature) in the most turbulent portions of the dataset.
Spectra from single tidal phases ($6\hrs$ slices of data), show that the high-wavenumber roll-off is not always detected.
In many cases, the spectrum is flat up to almost the highest wavenumber, where the spectrum often bends upwards.
This upward bend of the spectrum is due to the signal approaching the noise level of the measuring device in weaker stratification and less turbulent regions.

In the upper half of the mooring, the spectra do not reach the theoretical $-5/3$ slope, i.e.~they slope downwards to the right in figure \ref{fig:spectra}.
We suggest that this is the result of averaging together data regions having inertial ranges spanning different scales.
This effect is particularly strong during the downslope phase, being less turbulent than the upslope one.
We conclude that the thermistors resolve at least the lower-wavenumber part of the turbulence inertial range.
Turbulence is strongly intermittent, in particular during the downslope phase of the tide, as judged from the fact that the spectral slope is steeper than $-5/3$.

Similar conclusions could be drawn from the frequency spectra (avoiding the use of Taylor's hypothesis).
However, the spatial data have spectral slopes closer to the classical value of $-5/3$ during the upslope phase, and also better highlight the differences between the two tidal phases.

\subsection{Skewness and flatness of fluctuations}
\label{sec:fluctuations}

Figure \ref{fig:sk_ku_fluctuations} shows the skewness and flatness of temperature fluctuations, $\theta$, from the full dataset in the time domain, as a function of the height above the bottom and of the tidal phase.
The skewness is always negative and small (figure \ref{fig:sk_ku_fluctuations}, left panel).
The most negative skewness appears in the region above the near-bottom cold layer, most likely in connection with mixing of the cold layer itself; it decays towards zero further above.
The flatness (right panel in figure \ref{fig:sk_ku_fluctuations}) is close to the Gaussian value of 3, in particular in the upper half of the mooring and during the upslope phase of the tide.
The Gaussian value of the flatness is clearly exceeded only close to the bottom during the downslope phase.

As discussed in section \ref{sec:stats}, filtering the data is a naive attempt to separate turbulence from wave motions.
Skewness of the high-pass-filtered data (figure \ref{fig:sk_ku_fluctuations_filt}, left panel) is positive close to the bottom, while it is negative further above.
Deviations from zero are larger than in the unfiltered data, as expected for a turbulent signal where nonlinear dynamics dominate.
Moving towards the upper end of the mooring, skewness slowly approaches zero.
After filtering, skewness is similar during the two tidal phases, indicating that the differences observed in figure \ref{fig:sk_ku_fluctuations} can mostly be ascribed to internal wave motions.
The change of sign of skewness above the bottom is the signature of the overturns that mix the interface above the cold near-bottom layer.
Skewness has slightly larger negative values during the upslope phase.

The flatness of the filtered dataset (figure \ref{fig:sk_ku_fluctuations_filt}, right panel) is much larger than in the full data; it is also larger during the downslope phase.
Flatness increases towards the bottom, in particular in the lowest $10\m$ of the mooring.
A flatness larger that the Gaussian value of 3, with exponential tails of the pdf as the ones found in our data (not shown), is a common feature of turbulent fluctuations in the inertial and dissipation ranges of both active scalars \citep{ching_probabilities_1991} and passive scalars in the presence of a mean gradient \citep{shraiman_scalar_2000,warhaft_passive_2000}.

From the comparison of filtered and unfiltered data, we conclude that even if the internal waves dominating the total signal seem highly nonlinear and continuously breaking, as suggested by figure \ref{fig:temperature}, their statistics are not too far from Gaussian.
In the filtered data, on the other hand, turbulent fluctuations are strongly non-Gaussian.
Furthermore, the skewness and flatness of the wave motions are affected by the phase of the tide more than the skewness and flatness of the turbulent motions.
Both filtered and unfiltered data depend on the height above the bottom, in particular in segment A.

\subsection{Skewness and flatness of increments}
\label{sec:sk_and_fl}

We now consider the skewness and flatness of temperature increments, defined by equations \ref{eq:norm_mom} and \ref{eq:incrm}, plotted in figure \ref{fig:sk_fl_increments} as a function of the separation distance $r$.
Here, the values are presented as averages in the four segments.
Note that $\Dtr$ has a mean different from zero, since the tidal wave produces a trend in $\theta$; the mean increment is greater or smaller than zero depending on the tidal phase (downslope or upslope respectively).

Skewness has a rather complex dependence on $r$.
During both tidal phases, skewness increases from the bottom (A) to the top (D) of the mooring.
For $r$ smaller than  $5-10\m$, skewness is higher during the upslope phase compared to the downslope phase, being positive only for segments C and D during the upslope phase.
For separation distances larger than $10\m$, skewness changes in opposite directions for the two tidal phases, towards more positive values during the downslope phase and towards more negative values during the upslope phase.
This behaviour for larger scales is due to the opposite mean trend of temperature due to the tidal wave itself.

In the inertial range, a skewness of the temperature increments different from zero is considered the signature of ramp--cliff structures \citep[][]{warhaft_passive_2000}.
These ramp--cliff structures are most visible if increments are computed parallel to the mean scalar gradient (vertical), but they have a projection also on the horizontal plane.
Here we focus on horizontal increments, owing to the limitations of the vertical resolution.
The observed behaviour of skewness shows that the anisotropy of turbulence dynamics (as measured by skewness) depends on the phase of the tide and on the height above the bottom at the smallest scales resolved in the dataset.
If skewness is determined by small-scale convective plumes \citep[producing structures that can be assimilated to ramp--cliff ones,][]{zhou_plume_2002}, the observed differences in skewness could be caused by the dominance of either warm or cold convective plumes, depending on the tidal phases.
\citet{thorpe_skewness_1991} suggested that negative skewness is associated with convection, using observations from near-surface and near-bottom oceanic boundary layers.
Here, the clearest signs of convection are found in segments C and D during the upslope tide (as will be discussed in sections \ref{sec:saturation} and \ref{sec:plus_and_minus}), having positive skewness at small separations.
The difference between our results and those of \citet{thorpe_skewness_1991} could be caused by the prevalence here of warm rather than cold plumes, a hypothesis that is discussed in more detail below.
More generally, we suggest that the skewness of the temperature increments is not a good indicator of the presence or absence of convective activity.

Flatness decreases for increasing separation distances during both tidal phases, approaching without actually reaching the Gaussian value of 3 at the largest scales considered.
A broad range of power-law decay is evident, as often observed within the inertial range in the laboratory \citep[e.g.~figure 3 in][]{warhaft_passive_2000}.
Flatness is expected to saturate for $r$ comparable to the dissipation scale, but we barely see a hint of such a saturation at the smallest $r$, confirming that the dissipation scale is not resolved here.
Flatness is very similar throughout the mooring during the downslope phase, while it is higher away from the bottom during the upslope phase.

In general, skewness and flatness strongly depend on $r$.
Skewness in particular has a rich behaviour within the turbulence inertial range identified in the spectra in figure \ref{fig:spectra}.
However, variations in skewness are generally smooth, and overall these quantities do not provide any clear suggestion of a ``scale break'' between turbulence and wave dynamics.
In the next section, we will show that, on the other hand, higher-order moments give a clear indication of such a scale break.

\subsection{High-order moments}
\label{sec:saturation}

We extend the description of the statistics of the temperature increments to higher-order moments, using in particular the generalised structure function defined in \ref{eq:gsf}.
In order to measure the convergence of the computation of these higher-order moments, we consider the quantity
\begin{equation}
    \frac{\max{\left|\Dtr\right|^q}}{M\,\gamma_q(r)},
    \label{eq:conv}
\end{equation}
where $M$ is the number of data points used to compute the average $\gamma_q$.
This ratio is the relative contribution of the largest resolved increment (at scale $r$) to the generalised structure function.
A well-converged generalised structure function implies a small value of this ratio, since the increments most relevant for the order $q$ are then well resolved, i.e.~they are smaller than $\max{\left|\Dtr\right|}$.
In order to improve the convergence of $\gamma_q$ for higher $q$, we consider the statistics averaged in two vertical subsets of the mooring instead of four as done in the previous sections.
We consider the lower half (segments A and B combined together) and the upper half (segments C and D combined together).

Based on the ratio \ref{eq:conv}, $\gamma_q$ is well converged at scales larger than $1\m$ at all orders (ratio $\mathcal{O}(10^{-2})$), with the partial exception of the highest orders in the lower half of the mooring during the upslope phase (not shown).
For scales smaller than $1\m$, the convergence is worse in particular during the upslope phase, and the orders higher than $q\approx 6$ are unreliable.
For this reason, we will focus on scales larger than $1\m$.

Figure \ref{fig:Rm} shows, in log--log space, the values of $\gamma_q$ as a function of $r$, for $q$ ranging from 1 to 10, in the lower and upper halves of the mooring and during the two tidal phases separately.
The scaling exponents $\zeta_q$ (equation \ref{eq:zeta}) are estimated in the turbulence scaling range and are shown in figure \ref{fig:slopes} as a function of the order $q$.
Considering the lower half (A+B panels in figure \ref{fig:Rm}), $\gamma_2$ has a scaling exponent close to $2/3$ over almost two decades during the upslope phase.
A scaling range with exponent $1.4$ (dotted lines) up to $r\approx 100\m$ is approached for high $q$ in the case reported in the lower left panel of figure \ref{fig:Rm}.
Figure \ref{fig:slopes}, however, shows that the scaling exponents do not saturate at this value, but keep increasing with $q$.
During the downslope phase (figure \ref{fig:Rm}, lower right panel), $\gamma_2$ has a scaling exponent higher than $2/3$, and $\gamma_q$ has an exponent larger than $1.4$ at the highest orders computed.
A clear break of the scaling behaviour is visible at $r\approx 100\m$ for $q>5$, in particular during the downslope phase.
It is unlikely that this is a consequence of the violation of Taylor's hypothesis, because the scale of the break is smaller during the upslope phase, having larger mean velocity.
We would expect on the other hand the scale of the break to be proportional to the mean velocity, if it were caused by a breakdown of Taylor's hypothesis.
Higher-order moments thus identify two separate ranges, which could be interpreted as turbulence-dominated for $r$ smaller than approximately $100\m$ and wave-dominated for larger scales.

Considering now the upper half of the mooring (C+D panels in figure \ref{fig:Rm}), $\gamma_2$ has a scaling exponent greater than $2/3$, markedly so during the downslope phase.
The scaling behaviour at high $q$ extends up to scales not as large as in segments A+B ($30$--$50\m$).
Wave motions may thus reach smaller scales in the upper half of the mooring, and turbulent motions may be limited to a narrower range.
The value of $\zeta_q$ for the highest $q$'s is well below $1.4$ during the upslope phase (see figure \ref{fig:slopes}), while it is higher than $1.4$ during the downslope one.
Considering that $\gamma_2$ has a slope in excess of $2/3$, we hypothesise that also the slopes for higher $q$ are similarly biased with respect to controlled environment results.
If this is actually the case, the asymptotic exponent during the upslope phase would in fact approach the value of 0.8 measured for an active scalar in convective turbulence \citep{zhou_plume_2002}.

Considering the scaling exponents in figure \ref{fig:slopes} more generally, we see strong deviations from the non-intermittent $\zeta_q=q/3$ prediction, marked by the dashed line in the figure.
The strongest deviations are observed during the upslope tidal phase, in particular in the upper half of the mooring (C+D).
Saturation of the moments is clearly visible in this part of the dataset, already for $q=5$.
These higher-order moments thus show that the tidal phase actually has a strong impact on intermittency, and that the turbulence behaviour is far from universal in the dataset.

\subsection{Plus and minus increments}
\label{sec:plus_and_minus}

In the previous section, we suggested that the behaviour of temperature is similar to that of a passive scalar in the lower half of the mooring, while in the higher part the statistics deviate towards those of active scalars, in particular during the upslope phase.
To further investigate this issue, we consider the skewness of plus and minus increments, introduced in section \ref{sec:stats}.
The skewness of the distributions of $\Dtr^\pm$ and $\Dtz^\pm$ are shown in figure \ref{fig:sk_increments_pm}, as a function of the separation distance and for the four segments of the mooring separately.

During the downslope phase, the skewness of the vertical minus increments, $\muq{\Dtz^-}{3}$ (large, filled red triangles), is greater than $\muq{\Dtz^+}{3}$ (large, filled red squares) in all segments.
During the upslope phase of the tide this is also true, but the difference is much smaller in the lower segments that in  upper ones.
Considering the horizontal increments during the downslope phase of the tide, the skewness of the minus increments (small, red filled triangles) is larger at all scales and throughout the mooring.
In the upslope phase closer to the bottom (segments A and B), the skewness of plus and minus horizontal increments are the same, or close, with the exception of larger scales.
On the other hand, plus skewness (small, blue open squares) is larger in segments C and D at all scales.

The difference in skewness between plus and minus increments is of order 1, consistent with that reported in the laboratory by \citet{zhou_plume_2002}.
On the other hand, laboratory data show that $\muq{\mathrm{\Delta}_\tau\theta^+}{3}$ and $\muq{\mathrm{\Delta}_\tau\theta^-}{3}$ converge to the same value for large $\tau$.
This does not happen here, where an increase of the difference is observed for $r$ larger than $10$--$100\m$.
We should not conclude from this that plumes of all sizes are present, but rather that wave motions are also associated with asymmetric temperature gradients.
We also cannot rule out that contamination between the two tidal phases and the use of the Taylor's hypothesis may play a role at the largest scales.

These results are discussed with the help of the sketch in figure \ref{fig:plumes}.
We consider two ``plumes'', which we regard here generically as an asymmetric temperature anomaly, with a sharp leading front and a smoother tail.
In the upper part of the figure, the isotherms of two of these plumes are sketched in the spatial domain: (1) on the left a hot plume with an upper sharper front; and (2) on the right, a cold plume with a lower sharper front.
Two transects are taken through each of the two plumes, one vertical (green dashed arrow) and one horizontal (red continuous arrow).
The resulting temperature transects are sketched in the two panels below, with green dashed and red continuous lines for vertical and horizontal directions, respectively.
Consider the hot plume on the left.
Its leading front is sharper than its tail; the vertical transect thus shows a slower increase of the temperature and a faster decrease to the mean value.
This structure of the plume shows up as larger values of skewness for minus rather than plus vertical increments, up to separations comparable with the size of the plume.
The same reasoning applies to the horizontal direction and for a plume with an inclination to the mean flow given in the figure.
The result is a larger skewness for plus than for minus horizontal increments.
Note that a cold plume with opposite tilt would produce different results for the vertical increments, so the asymmetry is characterised unambiguously.
This is consistent with the estimates in segments C and D during the upslope tidal phase.
A similar reasoning can be applied to the cold plume in figure \ref{fig:plumes}, suggesting that in this case $\muq{\Dtz^+}{3}<\muq{\Dtz^-}{3}$ and $\muq{\Dtr^+}{3}<\muq{\Dtr^-}{3}$.

We can now interpret the results from figure \ref{fig:sk_increments_pm} as follows.
During the upslope phase, warm plume-like temperature anomalies are dominant in segments C and D, well above the bottom.
During the downslope phase, cold plume-like anomalies are instead dominant throughout the mooring.
The asymmetry of temperature anomalies has both a vertical and a horizontal component.
In other words, plumes also have a preferential horizontal orientation, pointing towards the direction of the mean flow, as shown in figure \ref{fig:plumes}.
On the other hand, in segments A and B during the upslope tide, temperature anomalies are more symmetric, in particular along the horizontal direction.
Overall, we do not have any clear indication of the size of the plumes.
We can only speculate that their typical size is at least of $\mathcal{O}(1\m)$, i.e.~at least comparable to the values of $r$ for which skewness departs from its saturated, small-scale value.
The large plume of figure \ref{fig:temperature} is thus an extreme case, and scanning through the temperature record confirms this.

\section{Discussion and conclusions}
\label{sec:conclusions}

We described the temperature statistics using observations from a moored thermistor array, deployed in the deep ocean above a sloping bottom.
The dynamics at the mooring location are strongly affected by the semidiurnal tidal wave, and we thus considered separately the statistics for two tidal phases, cooling (upslope) and warming (downslope).
The thermistor array does not reach the bottom, with the lowest sensor approximately $5\m$ above the bottom, and a bottom mixed layer, if present, is not visible in the data.
The results are one of the few studies on the high-order scalar statistics in stably, strongly stratified fluids.

\begin{figure}
  \centering
    \includegraphics[width=0.49\textwidth]{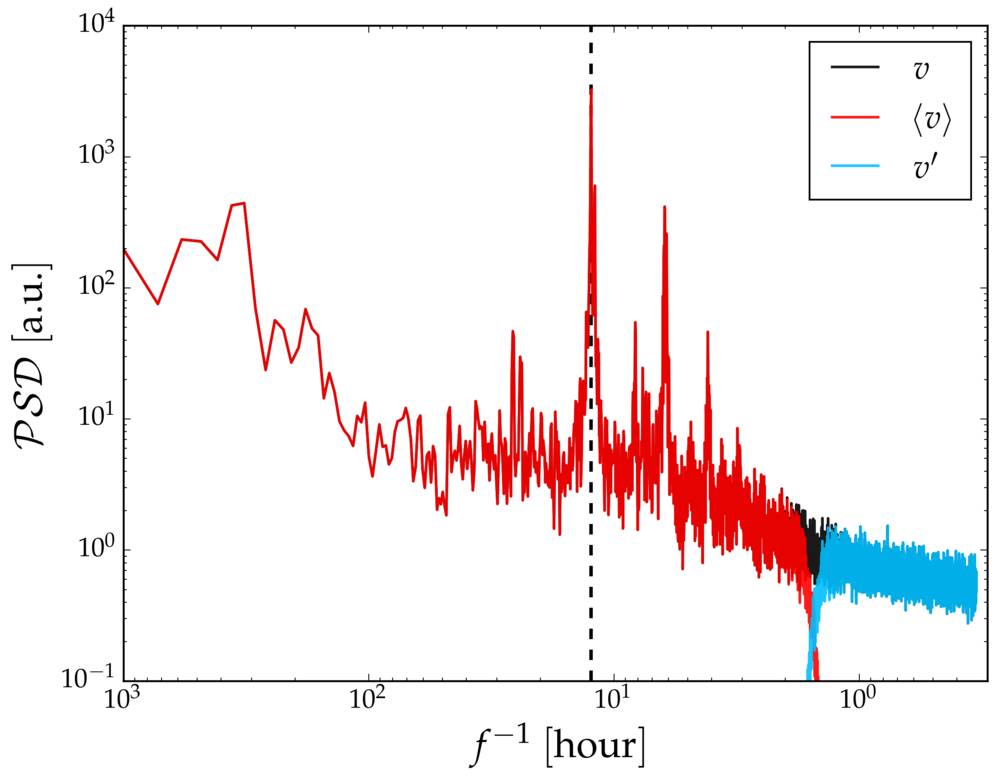}
    \includegraphics[width=0.49\textwidth]{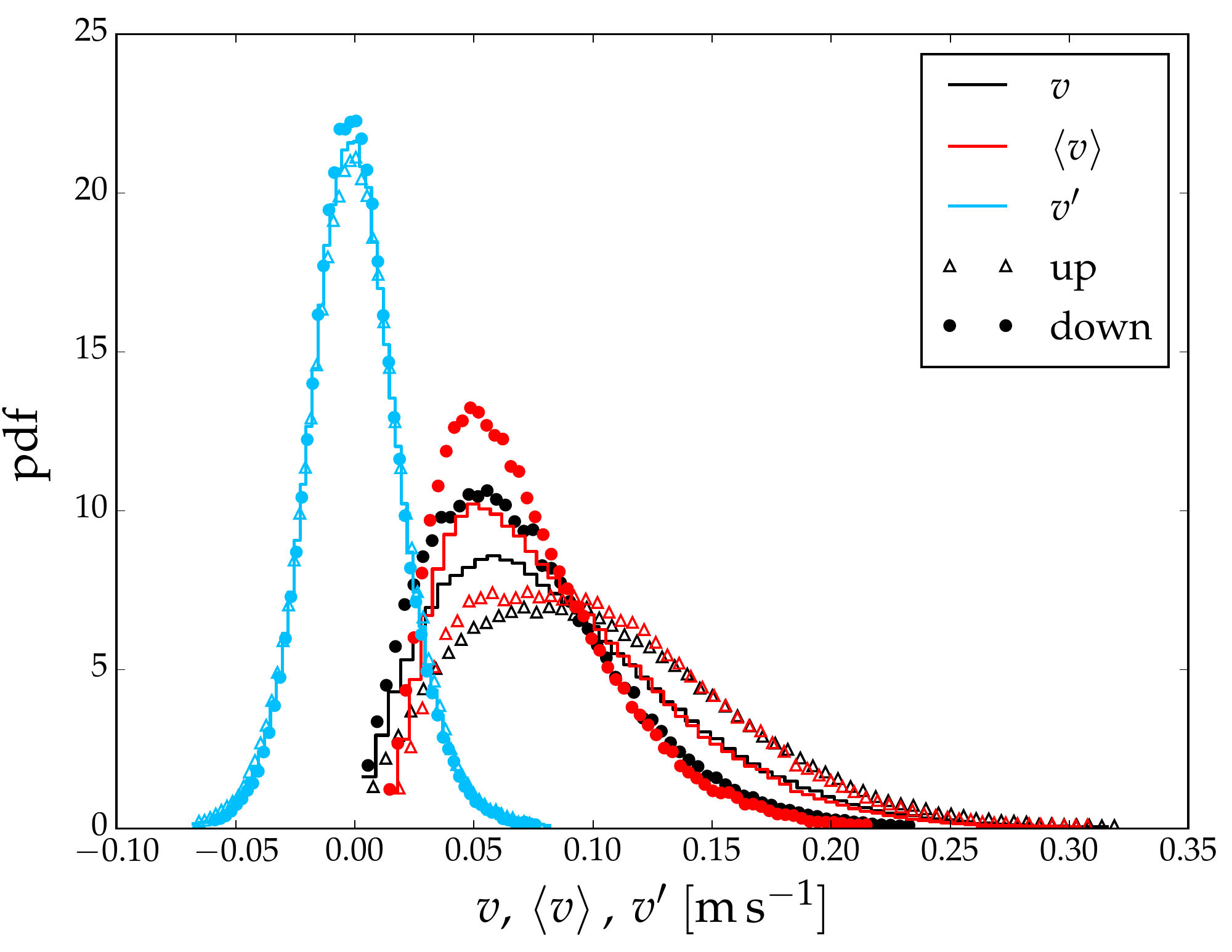}
    \caption{\label{fig:velocity}
      Spectra (left) and pdf's (right) of total velocity $v$ (black), mean velocity $\mean{v}$ (red, see text) and velocity fluctuations $v'$ (light blue).
      In the left panel, the thick dashed line shows the $M_2$ semidiurnal frequency.
      On the right panel, the lines show the pdf for all the data, while triangles (circles) show the pdf computed for the upslope (downslope) tidal phase alone.
      Only velocities measured within the depth range of the thermistor array are used in the figure.
    }
\end{figure}

\begin{figure}
  \centering
    \includegraphics[width=0.49\textwidth]{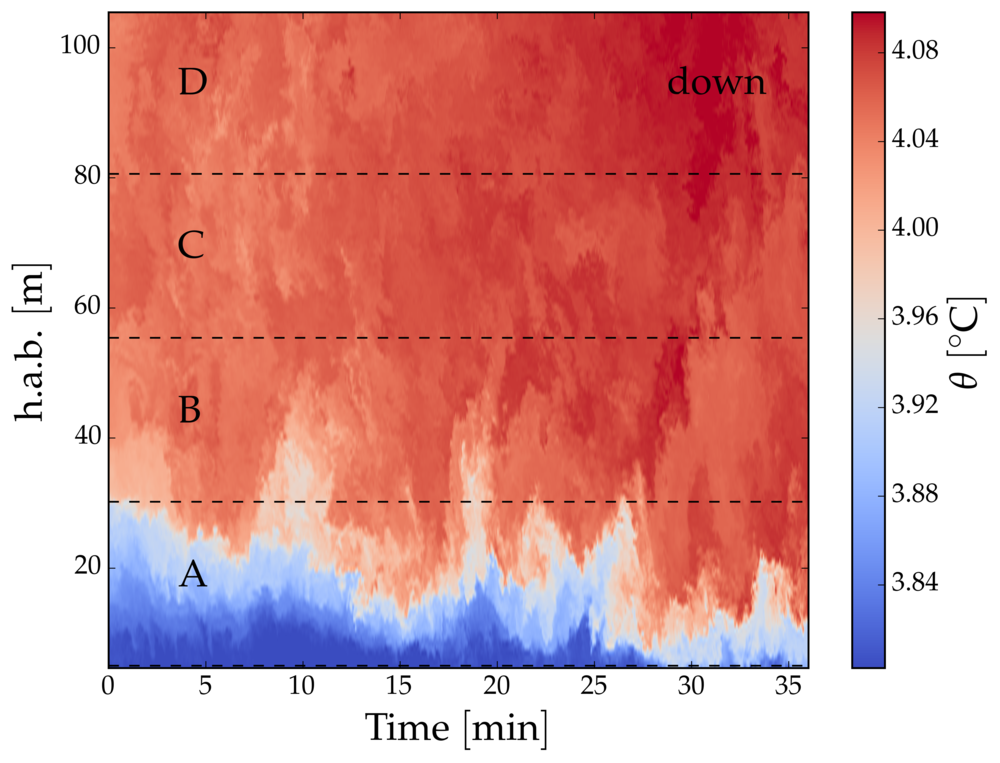}
    \includegraphics[width=0.49\textwidth]{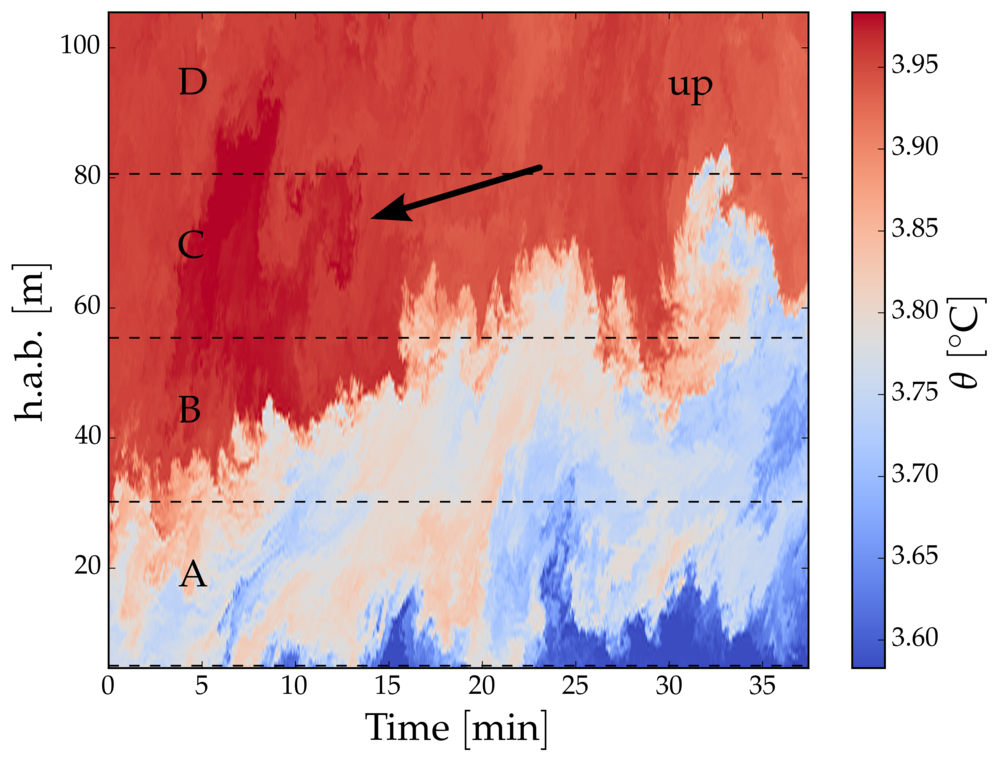}
    \caption{\label{fig:temperature}
      Examples from the dataset during a downslope (down) and upslope (up) tidal phase.
      The dashed lines mark the limits of the four segments of the mooring (A, B, C, D) considered separately throughout the paper.
      Colour scale is different in the two panels.
      The lowest thermistor is located approximately $5\m$ above the bottom.
      The vertical scale is in meters above the bottom.
      The black arrow points to the large plume mentioned in the text.
    }
\end{figure}

\begin{figure}
  \centering
    \includegraphics[width=0.7\textwidth]{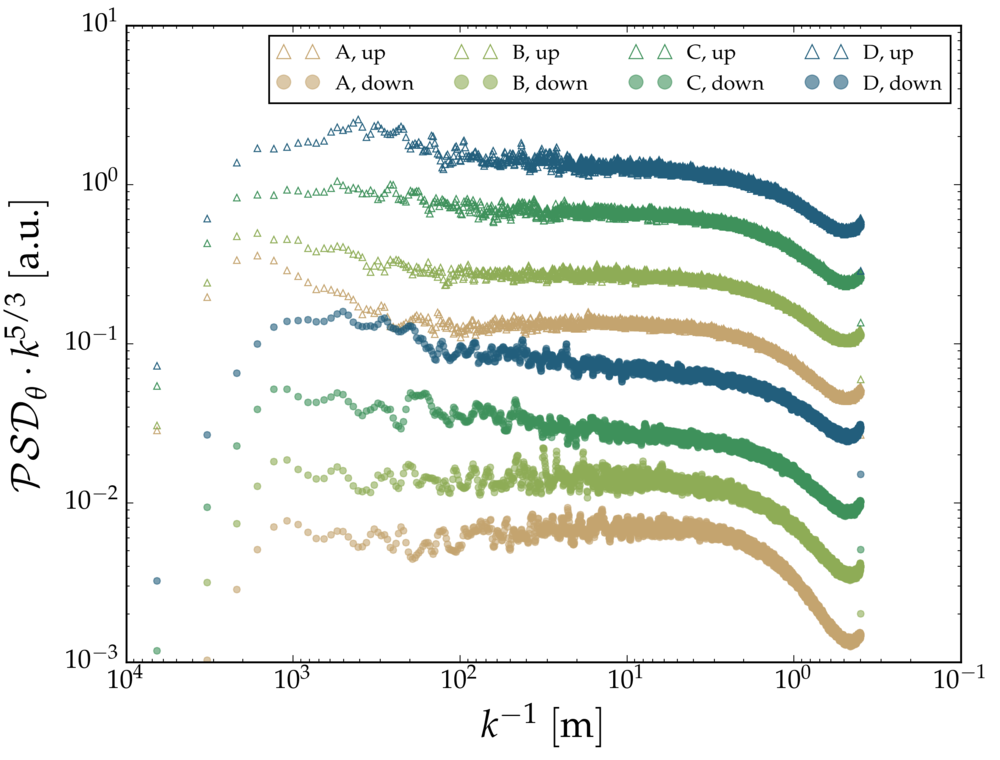}
    \caption{\label{fig:spectra}
        Average wavenumber spectra of temperature fluctuations, during the two tidal phases and in the four segments of the mooring, compensated by $k^{-5/3}$.
      An arbitrary vertical shift is applied to the spectra to better distinguish them.
    }
\end{figure}

\begin{figure}
  \centering
    \includegraphics[width=0.4\textwidth]{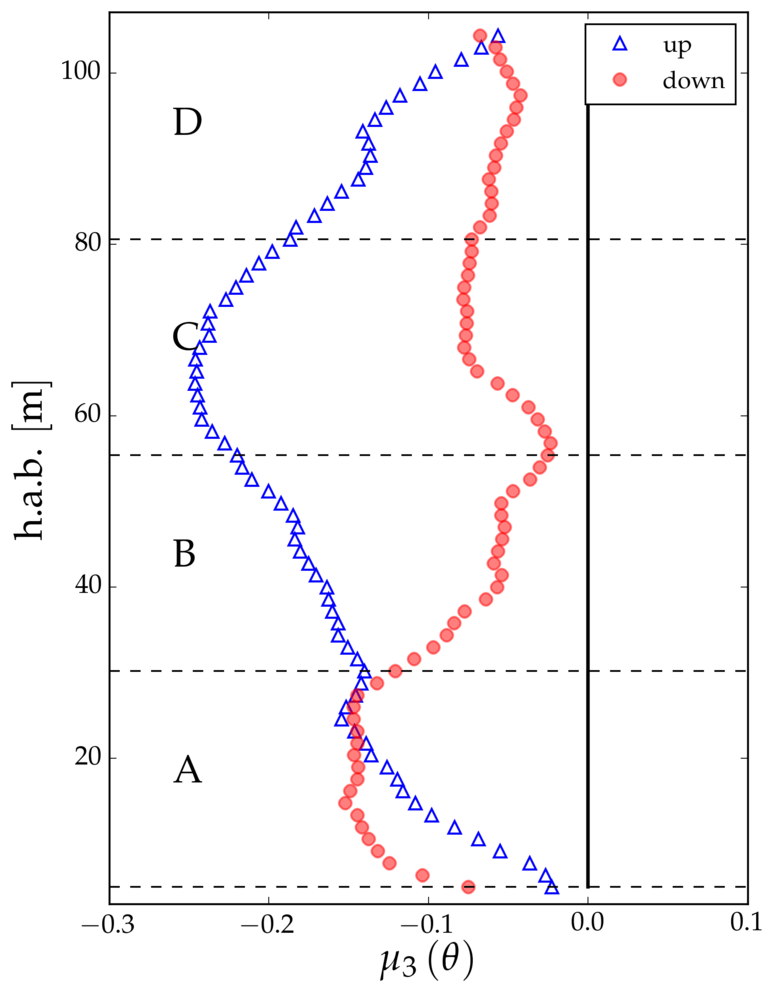}
    \includegraphics[width=0.4\textwidth]{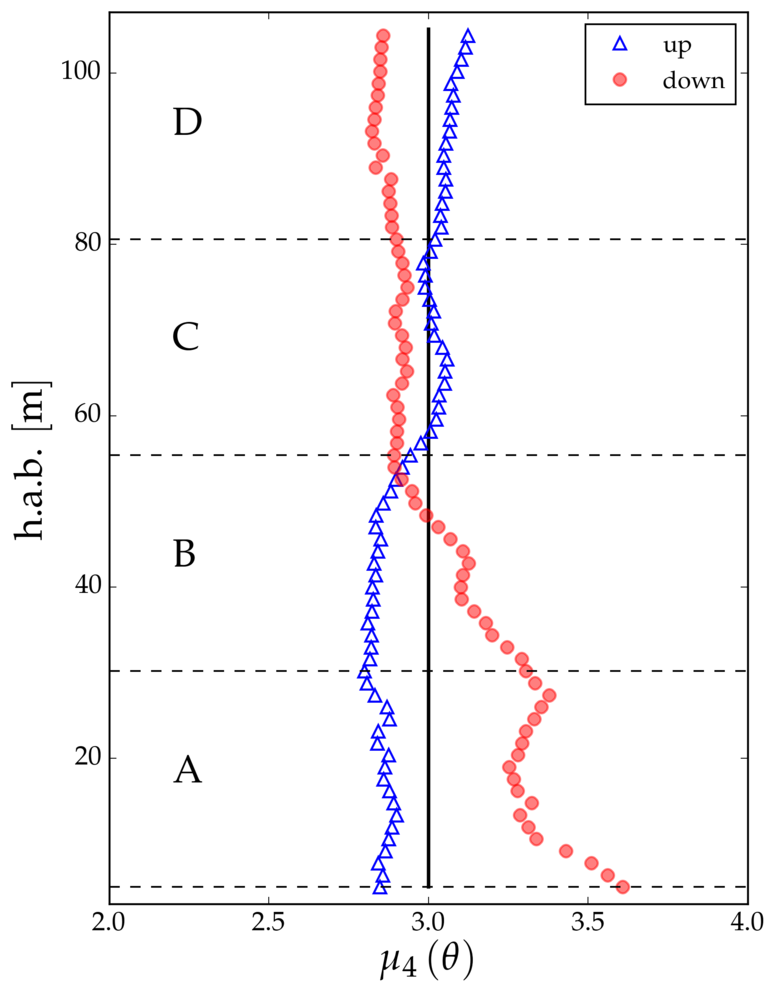}
    \caption{\label{fig:sk_ku_fluctuations}
      Skewness (left) and flatness (right) of temperature fluctuations, plotted as a function of depth for every other sensor in the string, and computed separately for different tidal phases.
      Results from the unfiltered time series (without using Taylor's hypothesis).
      The thick black lines mark the values of a Gaussian distribution.
      Dashed lines and letters refer to the different segments of the mooring (see table \ref{tab:segments}).
      Open blue triangles refer to the upslope phase, filled red circles refer to the downslope phase.
    }
\end{figure}

\begin{figure}
  \centering
    \includegraphics[width=0.4\textwidth]{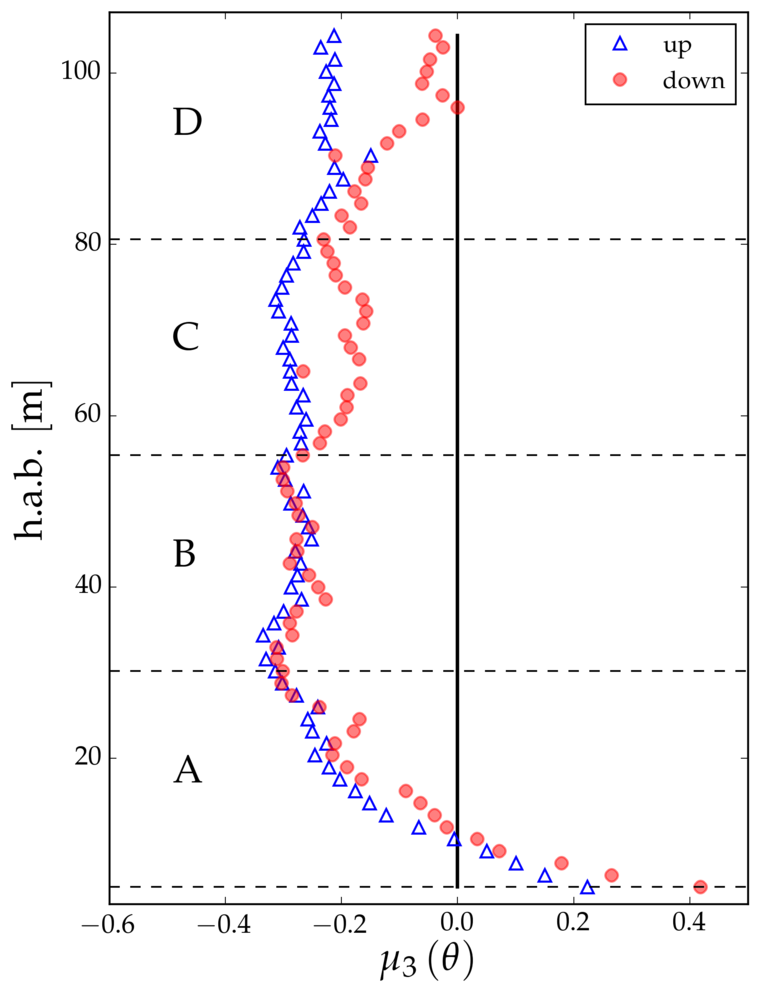}
    \includegraphics[width=0.4\textwidth]{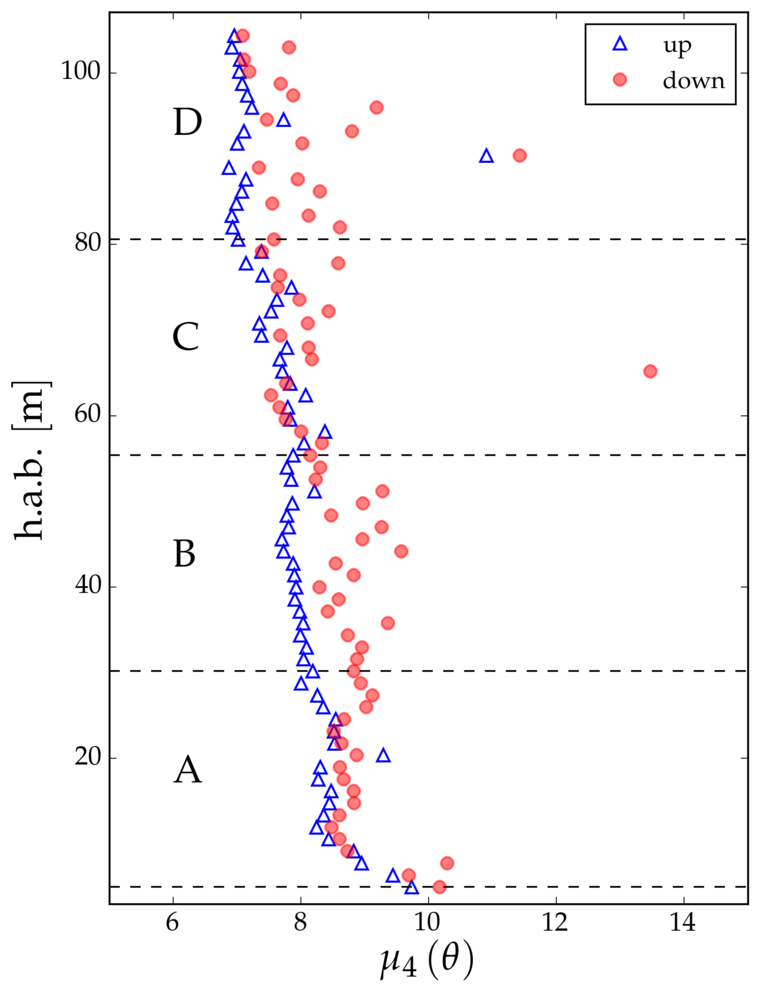}
    \caption{\label{fig:sk_ku_fluctuations_filt}
      As figure \ref{fig:sk_ku_fluctuations}, but for filtered data (see text).
      As a consequence of the filtering, the estimate of the flatness is noisier than in figure \ref{fig:sk_ku_fluctuations}.
    }
\end{figure}

\begin{figure}
  \centering
    \includegraphics[width=0.7\textwidth]{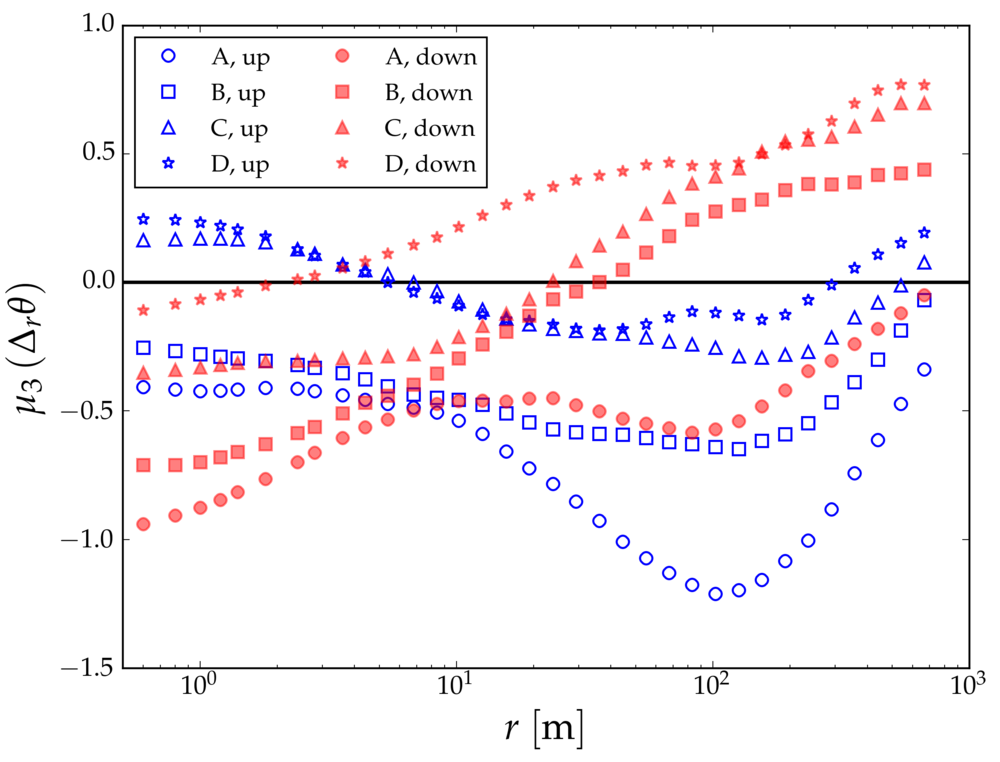}
    \includegraphics[width=0.7\textwidth]{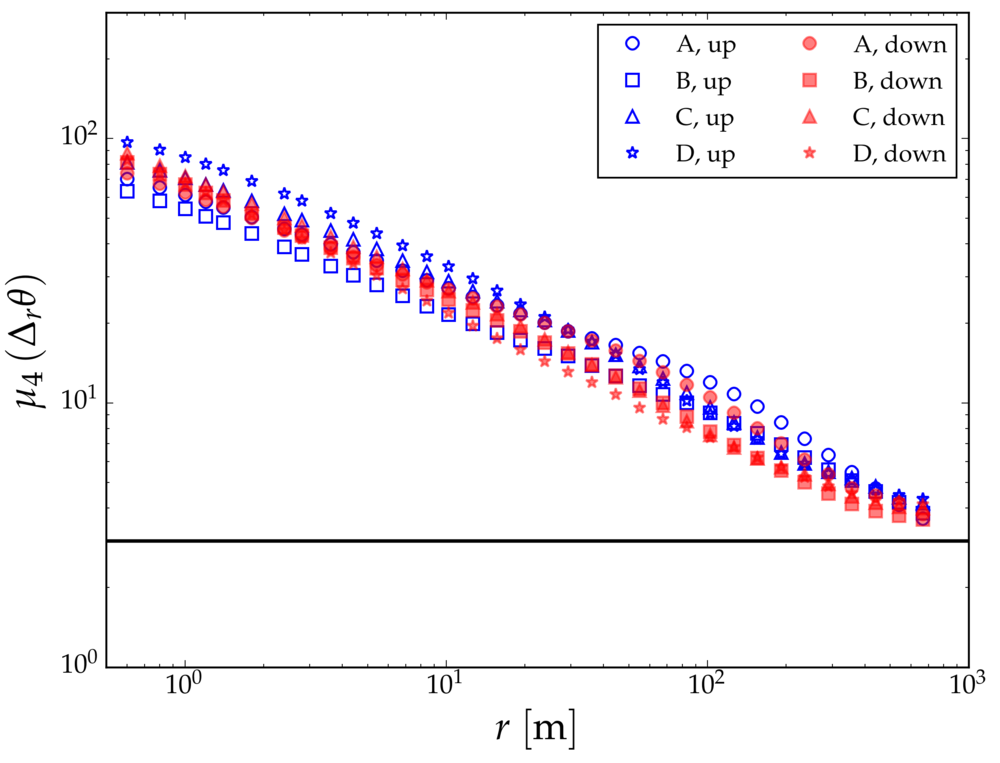}
    \caption{\label{fig:sk_fl_increments}
      Skewness ($\muq{\Dtr}{3}$) and flatness ($\muq{\Dtr}{4}$) of temperature increments for the four segments of the mooring, computed for upslope and downslope phase as a function of the separation $r$.
      The black lines mark the values for a Gaussian distribution.
      Note that the vertical axis of the right panel is logarithmic.
      Different colours refer to different tidal phases (open blue for upslope, filled red for downslope), different symbols to different segments of the mooring (see legend).
    }
\end{figure}

\begin{figure}
  \centering
    \includegraphics[width=\textwidth]{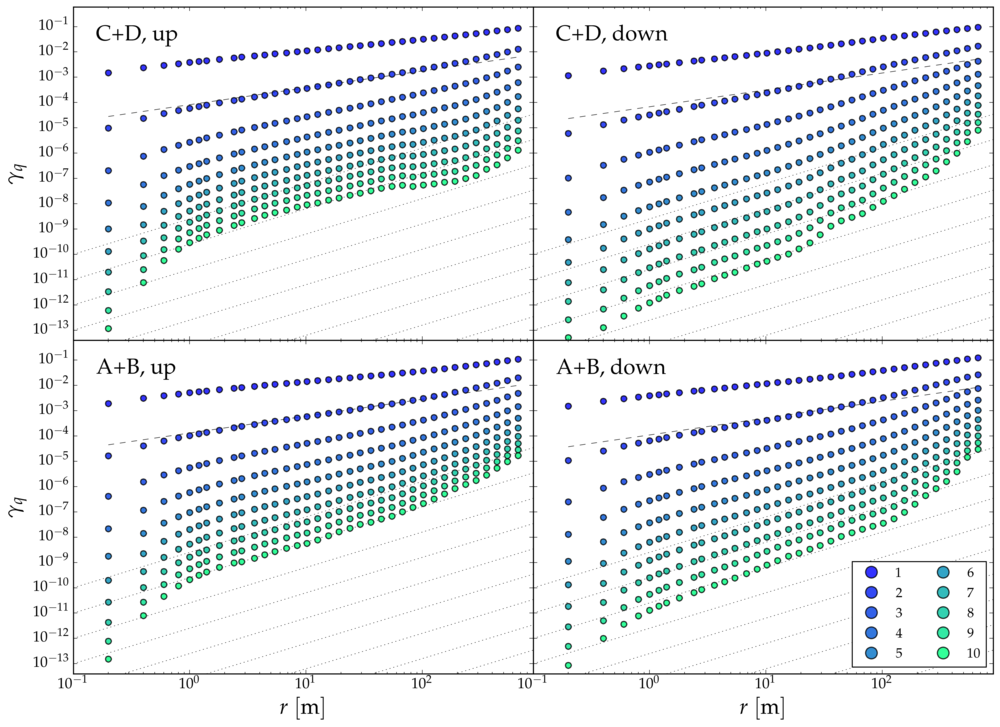}
    \caption{\label{fig:Rm}
      Generalised structure function of temperature increments $\gamma_q$.
      In the upper (lower) panels the results for the segments C, D (A, B) combined together are shown.
      The left panels refer to the upslope phase of the tide, while the right panels refer to the downslope phase.
      Different orders $q$ are plotted, increasing from top to bottom in each panel, with colours changing with $q$ (see legend).
      The black dashed line near the top is a guide for the eye with a slope of $2/3$, the theoretical slope for the structure function of order 2.
      The dotted black lines in the lower part of each panel have a slope of $1.4$, the asymptotic value of the slope in the inertial range for a passive scalar in grid turbulence \citep{warhaft_passive_2000}.
    }
\end{figure}

In the dataset, temperature fluctuations and temperature increments are strongly non-Gaussian at (horizontal) scales smaller than $10$--$100\m$.
At larger scales, deviations from Gaussian are smaller, but there generally is still a scaling behaviour, with a higher value of the scaling exponent $\zeta_q$.
The surprising result here is that high-order moments show an abrupt transition between the two regimes (see in particular panels A+B up and down, C+D down in figure \ref{fig:Rm}).
The length scale of the transition is of order $100\m$, but the actual value is state-dependent, with significant variations depending on the tidal phase and on the height above the bottom.
The $100\m$ scale is equal to the buoyancy length obtained as the ratio of the velocity scale ($10^{-1}\m\s^{-1}$; see figure \ref{fig:velocity}) and the buoyancy frequency ($10^{-3}\s^{-1}$, see table \ref{tab:stratification}), and is approximately 100 times larger than the Ozmidov scale \citep[estimated from the Thorpe scale, averaged over the whole dataset, computed in][]{cimatoribus_comparison_2014}.
Since the layer thickness scales with the buoyancy length in a stratified turbulent environment \citep{billant_self-similarity_2001}, we hypothesise that motions at larger horizontal scales are less nonlinear, i.e.~more wave-like, as overturns are limited by the layer thickness.
In this view, we argue that this break in the scaling behaviour could be a characteristic feature of stratified turbulence.
The fact that a sharp break is visible only in the high-order moments, at least in this dataset, could result from the fact that the buoyancy length is relevant only for turbulent motions.
On the other hand, internal waves are scale-free, at least according to the dispersion relation from linear theory, and thus have scales both smaller and larger than the break.
Since waves are less effective than turbulent motions at producing sharp gradients, their signature on higher-order moments is smaller than for turbulent fluctuations, and thus smooth the transition between the two regimes only in the lower-order moments.
The possibility to define a (state-dependent) scale separation between waves and turbulence may have important practical applications, e.g.~for large eddy simulations \citep{frola_experimental_2014}.
It is also noteworthy that, assuming we can indeed identify large-scale motions with waves, deviations from Gaussian are small even if there is ample evidence of nonlinearity and breaking.

We also considered generalised structure functions and skewness of plus/minus increments with respect to the passive/active nature of temperature dynamics.
There is a clear indication of a passive scalar behaviour during the upslope phase close to the bottom; shear is strong enough to dominate the buoyancy forces.
On the other hand, in the upper half of the mooring, temperature statistics approach those of an active scalar during the upslope phase.
The difference in skewness of plus/minus increments is consistent with the presence of warm convective plumes.
The numerical simulations of \citet{slinn_modeling_2003} and \citet{gayen_boundary_2011,gayen_negative_2011} suggested that tidal motions above slopes can produce strong convection and mixing, in particular during the upslope tidal phase.
Our results provide support for their conclusions.

\begin{figure}
  \centering
    \includegraphics[width=0.7\textwidth]{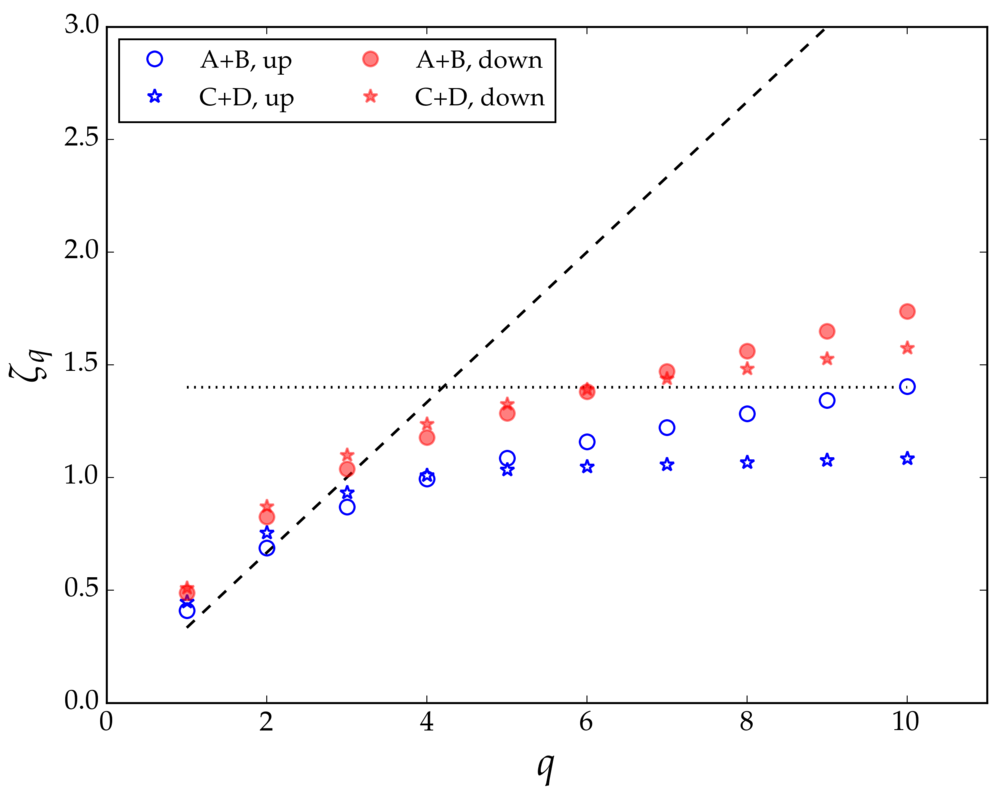}
    \caption{\label{fig:slopes}
      Exponents $\zeta_q$ of the generalised structure functions shown in figure \ref{fig:Rm} as a function of the order $q$, estimated by a least square fit in the scaling range of turbulence.
      The dashed line is the $\zeta_q=q/3$ slope for the non-intermittent case.
      The dotted horizontal line marks $\zeta=1.4$, the asymptotic value approached in grid turbulence for large $q$.
      Filled red (open blue) symbols refer to the downslope (upslope) tidal phase.
      Circles (stars) refer to the lower (upper) half of the mooring.
    }
\end{figure}

\begin{figure}
  \centering
    \includegraphics[width=\textwidth]{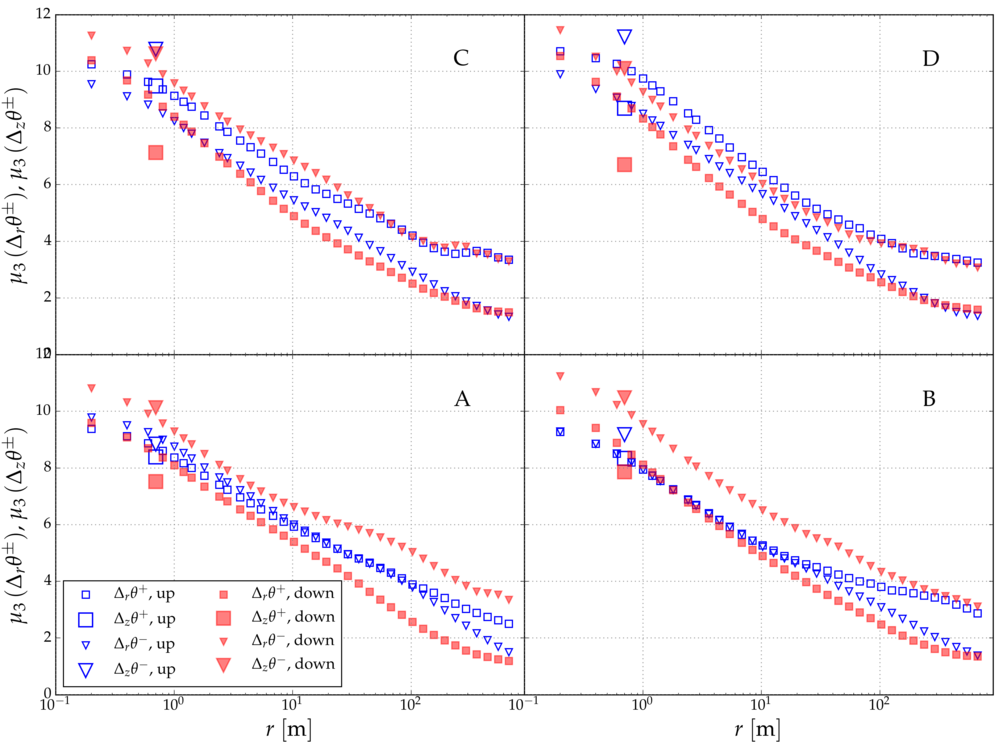}
    \caption{\label{fig:sk_increments_pm}
      Skewness of plus and minus increments as a function of the separation distance $r$.
      The four panels refer to the A--D segments of the mooring.
      Colours refer to different tidal phases (open blue for upslope, filled red for downslope), different symbols to plus and minus increments (squares for plus, triangles for minus).
      The skewness of both horizontal increments $\muq{\Dtr^\pm}{3}$ (small symbols) and vertical increments  $\muq{\Dtz^\pm}{3}$ (large symbols) are reported (see legend).
    }
\end{figure}

\begin{figure}
  \centering
    \includegraphics[width=0.8\textwidth]{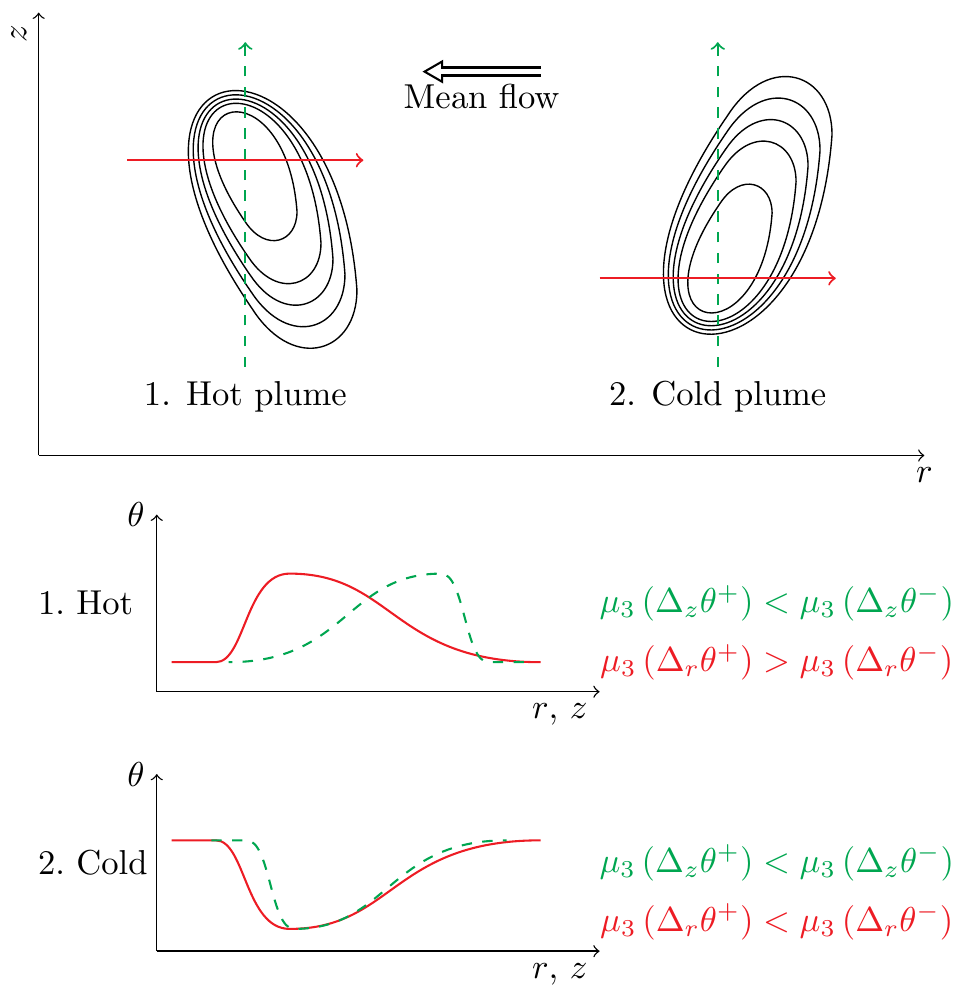}
    \caption{\label{fig:plumes}
      Sketch of the signature of convective plumes found in the dataset, see text for an explanation.
    }
\end{figure}

During the downslope phase, results are more ambiguous.
The difference in skewness of plus/minus increments indicates an asymmetry of temperature anomalies consistent with the presence of cold plumes throughout the mooring.
However, the scaling exponents of the generalised structure function exceed those of passive scalars (figure \ref{fig:slopes}).
The fact that in this case low-order moments exceed the $q/3$ prediction suggests, however, that this less turbulent tidal phase is strongly inhomogeneous, and a comparison with laboratory results is more difficult.
We hypothesise that the observed scaling behaviour is typical of strongly stratified, sporadically turbulent flows.
In this respect, it would be useful to compare our results with numerical simulations similar to those reported in \citet{rorai_turbulence_2014}, if scalar statistics can be computed from them.

In conclusion, we want to return to the initial, broader, issue of mixing in geophysical flows.
Overall, the indication from our results is that temperature statistics are highly dependent on the state of the flow, even if stratification is always present.
Differences in the statistics, in particular in relation to convective motions, are very likely to be connected to changes in mixing properties, in particular mixing efficiency.
A clearer understanding of mixing in a stably stratified flow is an important open question with implications in the fields of oceanography and climate.
We hope that the present results can help in this effort.\\

  The authors would like to thank the crew of RV~{\em Pelagia} for enabling the deployment and recovery of the instruments, and all the technicians, in particular Martin Laan, indispensable for the success of the measurements.
The authors would like to thank the referees for their useful and detailed comments on the previous versions of the manuscript.

%\bibliographystyle{apalike}
%\bibliography{Library_s}

\begin{thebibliography}{}

\bibitem[Alford and Pinkel, 2000]{alford_observations_2000}
Alford, M.~H. and Pinkel, R. (2000).
\newblock Observations of overturning in the thermocline: The context of ocean
  mixing.
\newblock {\em J.~Phys.~Oceanogr.}, 30:805--832.

\bibitem[Billant and Chomaz, 2001]{billant_self-similarity_2001}
Billant, P. and Chomaz, J.-M. (2001).
\newblock Self-similarity of strongly stratified inviscid flows.
\newblock {\em Phys.~Fluids}, 13:1645--1651.

\bibitem[Brethouwer and Lindborg, 2008]{brethouwer_passive_2008}
Brethouwer, G. and Lindborg, E. (2008).
\newblock Passive scalars in stratified turbulence.
\newblock {\em Geophys.~Res.~Lett.}, 35:L06809.

\bibitem[Celani et~al., 2001]{celani_fronts_2001}
Celani, A., Lanotte, A., Mazzino, A., and Vergassola, M. (2001).
\newblock Fronts in passive scalar turbulence.
\newblock {\em Phys.~Fluids}, 13:1768--1783.

\bibitem[Celani et~al., 2002]{celani_scaling_2002}
Celani, A., Matsumoto, T., Mazzino, A., and Vergassola, M. (2002).
\newblock Scaling and universality in turbulent convection.
\newblock {\em Phys. Rev. Lett.}, 88:054503.

\bibitem[Ching, 1991]{ching_probabilities_1991}
Ching, E.~S. (1991).
\newblock Probabilities for temperature differences in {Rayleigh}-{B\'enard}
  convection.
\newblock {\em Phys.~Rev.~A}, 44:3622--3629.

\bibitem[Cimatoribus et~al., 2014]{cimatoribus_comparison_2014}
Cimatoribus, A.~A., {van Haren}, H., and Gostiaux, L. (2014).
\newblock Comparison of {E}llison and {T}horpe scales from {E}ulerian ocean
  temperature observations.
\newblock {\em J.~Geophys.~Res.-Oceans}, 119:7047--7065.

\bibitem[{Costa Frola} et~al., 2014]{frola_experimental_2014}
{Costa Frola}, E., Mazzino, A., Cassola, F., Mortarini, L., and Ferrero, E.
  (2014).
\newblock An experimental study of the statistics of temperature fluctuations
  in the atmospheric boundary layer.
\newblock {\em Bound.-Lay.~Meteorol.}, 150:91--106.

\bibitem[Davis et~al., 1994]{davis_multifractal_1994}
Davis, A., Marshak, A., Wiscombe, W., and Cahalan, R. (1994).
\newblock Multifractal characterizations of nonstationarity and intermittency
  in geophysical fields: Observed, retrieved, or simulated.
\newblock {\em J.~Geophys.~Res.-Atmos.}, 99:8055--8072.

\bibitem[Frisch, 1996]{frisch_turbulence_1996}
Frisch, U. (1996).
\newblock {\em Turbulence}.
\newblock Cambridge University Press, Cambridge, {U.K.}

\bibitem[Garrett, 1990]{garrett_role_1990}
Garrett, C. (1990).
\newblock The role of secondary circulation in boundary mixing.
\newblock {\em J.~Geophys.~Res.-Oceans}, 95:3181--3188.

\bibitem[Garrett, 1991]{garrett_marginal_1991}
Garrett, C. (1991).
\newblock Marginal mixing theories.
\newblock {\em Atmos. Ocean}, 29:313--339.

\bibitem[Gayen and Sarkar, 2011a]{gayen_boundary_2011}
Gayen, B. and Sarkar, S. (2011a).
\newblock Boundary mixing by density overturns in an internal tidal beam.
\newblock {\em Geophys.~Res.~Lett.}, 38:L14608.

\bibitem[Gayen and Sarkar, 2011b]{gayen_negative_2011}
Gayen, B. and Sarkar, S. (2011b).
\newblock Negative turbulent production during flow reversal in a stratified
  oscillating boundary layer on a sloping bottom.
\newblock {\em Phys.~Fluids}, 23:101703.

\bibitem[Lamb, 2014]{lamb_internal_2014}
Lamb, K.~G. (2014).
\newblock Internal wave breaking and dissipation mechanisms on the continental
  slope/shelf.
\newblock {\em Ann.~Rev.~Fluid Mech.}, 46:231--254.

\bibitem[Lindborg and Fedina, 2009]{lindborg_vertical_2009}
Lindborg, E. and Fedina, E. (2009).
\newblock Vertical turbulent diffusion in stably stratified flows.
\newblock {\em Geophys.~Res.~Lett.}, 36:L01605.

\bibitem[Moum et~al., 2004]{moum_convectively_2004}
Moum, J.~N., Perlin, A., Klymak, J.~M., Levine, M.~D., Boyd, T., and Kosro,
  P.~M. (2004).
\newblock Convectively driven mixing in the bottom boundary layer.
\newblock {\em J.~Phys.~Oceanogr.}, 34:2189--2202.

\bibitem[Munk and Wunsch, 1998]{munk_abyssal_1998}
Munk, W. and Wunsch, C. (1998).
\newblock Abyssal recipes {II}: energetics of tidal and wind mixing.
\newblock {\em Deep-Sea Res. Pt I}, 45:1977--2010.

\bibitem[Munk, 1966]{munk_abyssal_1966}
Munk, W.~H. (1966).
\newblock Abyssal recipes.
\newblock {\em Deep-Sea Res.}, 13:707--730.

\bibitem[Mydlarski and Warhaft, 1998]{mydlarski_passive_1998}
Mydlarski, L. and Warhaft, Z. (1998).
\newblock Passive scalar statistics in high-{P}\'eclet-number grid turbulence.
\newblock {\em J.~Fluid Mech.}, 358:135--175.

\bibitem[Osborn and Cox, 1972]{osborn_oceanic_1972}
Osborn, T.~R. and Cox, C.~S. (1972).
\newblock Oceanic fine structure.
\newblock {\em Geophys.~Fluid Dyn.}, 3:321--345.

\bibitem[Pinton and Labb\'e, 1994]{pinton_correction_1994}
Pinton, J.-F. and Labb\'e, R. (1994).
\newblock Correction to the {T}aylor hypothesis in swirling flows.
\newblock {\em J.~Physique {II}}, 4:1461--1468.

\bibitem[Riley and Lindborg, 2008]{riley_stratified_2008}
Riley, J.~J. and Lindborg, E. (2008).
\newblock Stratified turbulence: A possible interpretation of some geophysical
  turbulence measurements.
\newblock {\em J.~Atmos.~Sci.}, 65:2416--2424.

\bibitem[Rorai et~al., 2014]{rorai_turbulence_2014}
Rorai, C., Mininni, P.~D., and Pouquet, A. (2014).
\newblock Turbulence comes in bursts in stably stratified flows.
\newblock {\em Phys.~Rev.~E}, 89:043002.

\bibitem[Seuront et~al., 1999]{seuront_multifractal_1999}
Seuront, L., Schmitt, F., Schertzer, D., Lagadeuc, Y., and Lovejoy, S. (1999).
\newblock Multifractal intermittency of {E}ulerian and {L}agrangian turbulence
  of ocean temperature and plankton fields.
\newblock {\em Nonlin.~Processes~Geophys.}, 3:236--246.

\bibitem[Shraiman and Siggia, 2000]{shraiman_scalar_2000}
Shraiman, B.~I. and Siggia, E.~D. (2000).
\newblock Scalar turbulence.
\newblock {\em Nature}, 405:639--646.

\bibitem[Slinn and Levine, 2003]{slinn_modeling_2003}
Slinn, D.~N. and Levine, M.~D. (2003).
\newblock Modeling internal tides and mixing over ocean ridges.
\newblock In {\em Near-Boundary Processes and Their Parameterization:
  Proc.~13th ``{A}ha {H}uliko'' a {H}awaiian Winter Workshop}, pages 59--68.

\bibitem[Smyth and Moum, 2012]{smyth_ocean_2012}
Smyth, W. and Moum, J. (2012).
\newblock Ocean mixing by {K}elvin-{H}elmholtz instability.
\newblock {\em Oceanography}, 25:140--149.

\bibitem[Tennekes and Lumley, 1972]{tennekes_first_1972}
Tennekes, H. and Lumley, J.~L. (1972).
\newblock {\em A First Course in Turbulence}.
\newblock {MIT} Press, Cambridge, Massachusetts.

\bibitem[Thoroddsen and Van~Atta, 1992]{thoroddsen_exponential_1992}
Thoroddsen, S.~T. and Van~Atta, C.~W. (1992).
\newblock Exponential tails and skewness of density-gradient probability
  density functions in stably stratified turbulence.
\newblock {\em J.~Fluid Mech.}, 244:547--566.

\bibitem[Thorpe, 2012]{thorpe_kelvinhelmholtz_2012}
Thorpe, S.~A. (2012).
\newblock On the {Kelvin-Helmholtz} route to turbulence.
\newblock {\em J.~Fluid Mech.}, 708:1--4.

\bibitem[Thorpe et~al., 1991]{thorpe_skewness_1991}
Thorpe, S.~A., Cur\'e, M., and White, M. (1991).
\newblock The skewness of temperature derivatives in oceanic boundary layers.
\newblock {\em J.~Phys.~Oceanogr.}, 21:428--433.

\bibitem[Thorpe et~al., 1990]{thorpe_variability_1990}
Thorpe, S.~A., Hall, P., and White, M. (1990).
\newblock The variability of mixing at the continental slope.
\newblock {\em Philos.~T.~R.~Soc.~A}, 331(1616):183--194.

\bibitem[{van Haren}, 2013]{haren_stratified_2013}
{van Haren}, H. (2013).
\newblock Stratified turbulence and small-scale internal waves above deep-ocean
  topography.
\newblock {\em Phys.~Fluids}, 25:106604.

\bibitem[{van Haren} and Gostiaux, 2009]{van_haren_high-resolution_2009}
{van Haren}, H. and Gostiaux, L. (2009).
\newblock High-resolution open-ocean temperature spectra.
\newblock {\em J.~Geophys.~Res.-Oceans}, 114:C05005.

\bibitem[{van Haren} and Gostiaux, 2010]{van_haren_deep-ocean_2010}
{van Haren}, H. and Gostiaux, L. (2010).
\newblock A deep-ocean {K}elvin-{H}elmholtz billow train.
\newblock {\em Geophys.~Res.~Lett.}, 37(3):L03605.

\bibitem[{van Haren} and Gostiaux, 2012]{van_haren_detailed_2012}
{van Haren}, H. and Gostiaux, L. (2012).
\newblock Detailed internal wave mixing above a deep-ocean slope.
\newblock {\em J.~Mar.~Res.}, 70:173--197.

\bibitem[{van Haren} et~al., 2009]{van_haren_nioz3:_2009}
{van Haren}, H., Laan, M., Buijsman, D.-J., Gostiaux, L., Smit, M.~G., and
  Keijzer, E. (2009).
\newblock {NIOZ3:} independent temperature sensors sampling yearlong data at a
  rate of 1 {Hz}.
\newblock {\em {IEEE} Journal of Oceanic Engineering}, 34:315--322.

\bibitem[{van Haren} et~al., 1994]{van_haren_measurements_1994}
{van Haren}, H., Oakey, N., and Garrett, C. (1994).
\newblock Measurements of internal wave band eddy fluxes above a sloping
  bottom.
\newblock {\em J.~Mar.~Res.}, 52:909--946.

\bibitem[Warhaft, 2000]{warhaft_passive_2000}
Warhaft, Z. (2000).
\newblock Passive scalars in turbulent flows.
\newblock {\em Ann.~Rev.~Fluid Mech.}, 32:203--240.

\bibitem[Winters and D'Asaro, 1996]{winters_diascalar_1996}
Winters, K.~B. and D'Asaro, E.~A. (1996).
\newblock Diascalar flux and the rate of fluid mixing.
\newblock {\em J.~Fluid Mech.}, 317:179--193.

\bibitem[Zhai et~al., 2010]{zhai_significant_2010}
Zhai, X., Johnson, H.~L., and Marshall, D.~P. (2010).
\newblock Significant sink of ocean-eddy energy near western boundaries.
\newblock {\em Nat.~Geosci.}, 3:608--612.

\bibitem[Zhou and Xia, 2002]{zhou_plume_2002}
Zhou, S.-Q. and Xia, K.-Q. (2002).
\newblock Plume statistics in thermal turbulence: Mixing of an active scalar.
\newblock {\em Phys.~Rev.~Lett.}, 89:184502.

\end{thebibliography}

\clearpage

\onecolumn

\begin{table}
  \centering
  \begin{tabular}{l | l}
    Latitude & $36^\circ\, 58.885'\Nl$\\
    Longitude & $13^\circ\, 45.523'\Wl$\\
    Deepest thermistor depth& $2205\m$\\
    Deepest thermistor h.a.b.& $5\m$\\
    Bottom slope & $9.4^\circ$ \\
    Number of thermistors & 144\\
    Thermistor vertical spacing & $0.7\m$\\
    Total length of array & $100.1\m$\\
    Total length of cable & $205\m$\\
    Deployment & 13 Apr 2013\\
    Recovery & 12 Aug 2013\\
  \end{tabular}
  \caption[Mooring details]{Details of the mooring from which the data is obtained.
    h.a.b.~is the height above the bottom.
    \label{tab:mooring}}
\end{table}

\begin{table}
  \centering
  \begin{tabular}{l | c c c c}
    Segment name & A & B & C & D \\
    \hline
    h.a.b.~range & $5.0$--$30.0\m$ & $30.0$--$55.0\m$ & $55.0$--$80.1\m$ & $80.1$--$105.1\m$ \\
    Thermistor number & $1$--$36$ & $37$--$72$ & $73$--$108$ & $109$--$144$\\
  \end{tabular}
  \caption[Segments details]{Definition of the segments used in the analysis of the data.
    h.a.b.~is the height above the bottom.
    \label{tab:segments}}
\end{table}

\begin{table}
  \centering
  \begin{tabular}{l l | c c c c | c}
    Quantity      & Tidal phase & \multicolumn{5}{c}{Segments} \\
    \hline
    $N$           &      & A      & B      & C      & D      & all         \\
    \hline
    \hline
    Mean          & Both & $1.46$ & $1.23$ & $1.26$ & $1.33$ & $1.32$ \\
                  & Up   & $1.43$ & $1.28$ & $1.29$ & $1.33$ & $1.33$ \\
                  & Down & $1.49$ & $1.17$ & $1.22$ & $1.34$ & $1.30$ \\
    \hline
    Percentile 1  & Both & $0.21$ & $0.18$ & $0.19$ & $0.21$ & $0.19$ \\
                  & Up   & $0.21$ & $0.19$ & $0.19$ & $0.21$ & $0.20$ \\
                  & Down & $0.20$ & $0.17$ & $0.18$ & $0.21$ & $0.19$ \\
    \hline
    Percentile 99 & Both & $6.15$ & $5.12$ & $5.06$ & $5.10$ & $5.40$ \\
                  & Up   & $5.98$ & $5.42$ & $5.32$ & $5.17$ & $5.48$ \\
                  & Down & $6.29$ & $4.79$ & $4.79$ & $5.03$ & $5.31$ \\
  \end{tabular}
  \caption[Buoyancy frequency]{Mean, 1th and 99th percentiles of buoyancy frequency (from reordered $25\s$ average profiles of temperature) in the four different segments of the mooring and for the full dataset.
    The results are in units of $10^{-3}\s^{-1}$.
    \label{tab:stratification}}
\end{table}

\clearpage

\onecolumn

\end{document}